\documentclass[fleqn,usenatbib]{mnras}

\usepackage{newtxtext,newtxmath}

\usepackage[T1]{fontenc}
\usepackage{ae,aecompl}
\usepackage{tikz}
\usepackage{bm}
\usepackage{relsize}
\usetikzlibrary{positioning}
\usepackage{listings}
\usepackage{xcolor} 
\usepackage{verbatim}

\DeclareRobustCommand{\VAN}[3]{#2}
\let\VANthebibliography\thebibliography
\def\thebibliography{\DeclareRobustCommand{\VAN}[3]{##3}\VANthebibliography}

\usepackage{graphicx}	
\usepackage{amsmath}	
\usepackage{color}
\usepackage{tikz}

\definecolor{linkcolor}{rgb}{0,0,0.25}
\hypersetup{
colorlinks=true,        
linkcolor=blue,    
citecolor=blue,    
filecolor=blue,    
urlcolor=blue,      
draft=False,
}

\makeatletter
\renewcommand{\@printed}{}
\makeatother

\newcommand{\figurename}{Figure}
\newcommand{\tablename}{Table}
\newcommand{\eqnname}{Equation}
\newcommand{\secname}{Section}

\definecolor{darkgreen}{rgb}{0.0, 0.7, 0.0}

\definecolor{darkblue}{rgb}{0.0, 0., 0.7}

\newcommand{\teff}{\ensuremath{T_\mathrm{eff}}}
\newcommand{\logg}{\ensuremath{\log g}}
\newcommand{\xh}[1]{\ensuremath{[\mathrm{#1/H}]}}

\newcommand{\dex}{\ensuremath{\mathrm{dex}}}

\newcommand{\plx}{\ensuremath{\varpi}}
\newcommand{\mas}{\ensuremath{\mathrm{mas}}}
\newcommand{\uas}{\ensuremath{\mu\mathrm{as}}}
\newcommand{\muas}{\uas}
\newcommand{\pc}{\ensuremath{\mathrm{pc}}}
\newcommand{\kpc}{\ensuremath{\mathrm{kpc}}}
\newcommand{\kms}{\ensuremath{\mathrm{km\,s}^{-1}}}
\newcommand{\kmskpc}{\ensuremath{\mathrm{km\,s}^{-1}\mathrm{kpc}^{-1}}}
\newcommand{\masyr}{\ensuremath{\mathrm{mas\,yr}^{-1}}}

\newcommand{\gaia}{\emph{Gaia}}

\newcommand{\thisr}{8.23}
\newcommand{\thisrerr}{0.12}

\title[Galactic barycenter distance]{A direct measurement of the distance to the Galactic center using the kinematics of bar stars}

\author[Leung et al.]{
Henry W. Leung$^{1}$\thanks{E-mail: henrysky.leung@utoronto.ca},
Jo Bovy$^{1,2}$,
J.~Ted Mackereth$^{1,2,3}$, 
Jason~A.~S. Hunt$^{4}$,
Richard~R. Lane$^{5}$,
John~C. Wilson$^{6}$
\newauthor
\\
$^{1}$David A. Dunlap Department of Astronomy and Astrophysics, University of Toronto, 50 St. George Street, Toronto, Ontario, M5S 3H4, Canada\\
$^{2}$Dunlap Institute for Astronomy and Astrophysics, University of Toronto, 50 St. George Street, Toronto, Ontario, M5S 3H4, Canada\\
$^{3}$Canadian Institute for Theoretical Astrophysics, University of Toronto, 60 St George Street, Toronto, ON M5S 3H8, Canada\\
$^{4}$Center for Computational Astrophysics, Flatiron Institute, 162 5th Avenue, New York City, NY 10010, USA\\
$^{5}$Centro de Investigación en Astronomía, Universidad Bernardo O'Higgins, Avenida Viel 1497, Santiago, Chile\\
$^{6}$Astronomy Department, University of Virginia, Charlottesville, VA 22901, USA
}

\date{}

\pubyear{2021}

\begin{document}
\label{firstpage}
\pagerange{\pageref{firstpage}--\pageref{lastpage}}
\maketitle

\begin{abstract}
The distance to the Galactic center $R_0$ is a fundamental parameter for understanding the Milky Way, because all observations of our Galaxy are made from our heliocentric reference point. The uncertainty in $R_0$ limits our knowledge of many aspects of the Milky Way, including its total mass and the relative mass of its major components, and any orbital parameters of stars employed in chemo-dynamical analyses. While measurements of $R_0$ have been improving over a century, measurements in the past few years from a variety of methods still find a wide range of $R_0$ being somewhere within $8.0$ to $8.5\,\kpc$. The most precise measurements to date have to assume that Sgr A$^*$ is at rest at the Galactic center, which may not be the case. In this paper, we use maps of the kinematics of stars in the Galactic bar derived from APOGEE DR17 and \gaia\ EDR3 data augmented with spectro-photometric distances from the \texttt{astroNN} neural-network method. These maps clearly display the minimum in the rotational velocity $v_T$ and the quadrupolar signature in radial velocity $v_R$ expected for stars orbiting in a bar. From the minimum in $v_T$, we measure $R_0 = \thisr \pm \thisrerr\,\kpc$. We validate our measurement using realistic $N$-body simulations of the Milky Way. We further measure the pattern speed of the bar to be $\Omega_\mathrm{bar} = 40.08\pm1.78\,\kmskpc$. Because the bar forms out of the disk, its center is manifestly the barycenter of the bar+disc system and our measurement is therefore the most robust and accurate measurement of $R_0$ to date. 
\end{abstract}




\section{Introduction}\label{sec:intro}

Our understanding of the present chemo-dynamical state of the Milky Way has dramatically improved in the past few years by the exquisite astrometric precision of the \gaia\ data \citep{2021A&A...649A...1G}. The precision of astrometric data has leaped dramatically from the $\approx 1\,\mas$ for HIPPARCOS in the early 1990s \citep{1997A&A...323L..49P} to the $\approx 0.02\,\mas$ for \gaia\ EDR3 \citep{2021A&A...649A...1G}. This improved astrometry for a large sample of stars allows us to determine the kinematics of stars within a few kiloparsec and has led to such discoveries as the Gaia Sausage (e.g., \citealt{2018MNRAS.478..611B, 2018Natur.563...85H}) and kinematic substructures in the disc (e.g., \citealt{2018Natur.561..360A, 2018MNRAS.479L.108K}) and halo (e.g., \citealt{2020MNRAS.493.4978S, 2020ApJ...891L..19I}). As the precision of our \emph{heliocentric} observations improves, the precision requirement on the parameters describing the transformation to the \emph{Galactocentric} coordinate frame grows, because Milky Way physics occurs in the Galactocentric frame. Foremost among these parameters is the Sun's location within the Galaxy: \gaia\ has allowed the Sun's height above the local mid-plane to be measured to $0.3\,\mathrm{pc}$ precision (or $1.4\%$; \citealt{2019MNRAS.482.1417B}), however, the Sun's distance $R_0$ to the center of the Milky Way remains uncertain. We are, in particular, interested in the distance $R_0$ to the Milky Way's \emph{barycenter}. This is the relevant center for most dynamical and galaxy-evolution studies of the Milky Way, but it need not coincide with the region of highest stellar density or with the location of the Milky Way's supermassive black hole Sgr A$^*$ \citep{1986MNRAS.221.1023K}. Most past measurements have assumed that the barycenter coincides with one of these options.

Considerable efforts have been made to determine $R_0$ over the past century. One of the first attempts to estimate $R_0$ was performed by \citet{1918ApJ....48..154S}, where they showed that the distribution of globular clusters forms a spherical system with center approximately $13\,\kpc$ away in the direction of the Sagittarius constellation. They assumed that this center is the Galactic center, but this failed to account for the yet-to-be-discovered effect of interstellar extinction \citep{1930LicOB..14..154T}, leading to an overestimation of $R_0$. Since this inauspicious start, multiple other authors have estimated $R_0$ from the apparent center of the spatial distributions of globular clusters (e.g., $8.5\,\kpc$ from \citealt{1976AJ.....81.1095H}) and of stars (e.g., $7.7\,\kpc$ from Cepheid variables; \citealt{2009MNRAS.398..263M}).

\begin{figure}
\centering
\includegraphics[width=0.475\textwidth]{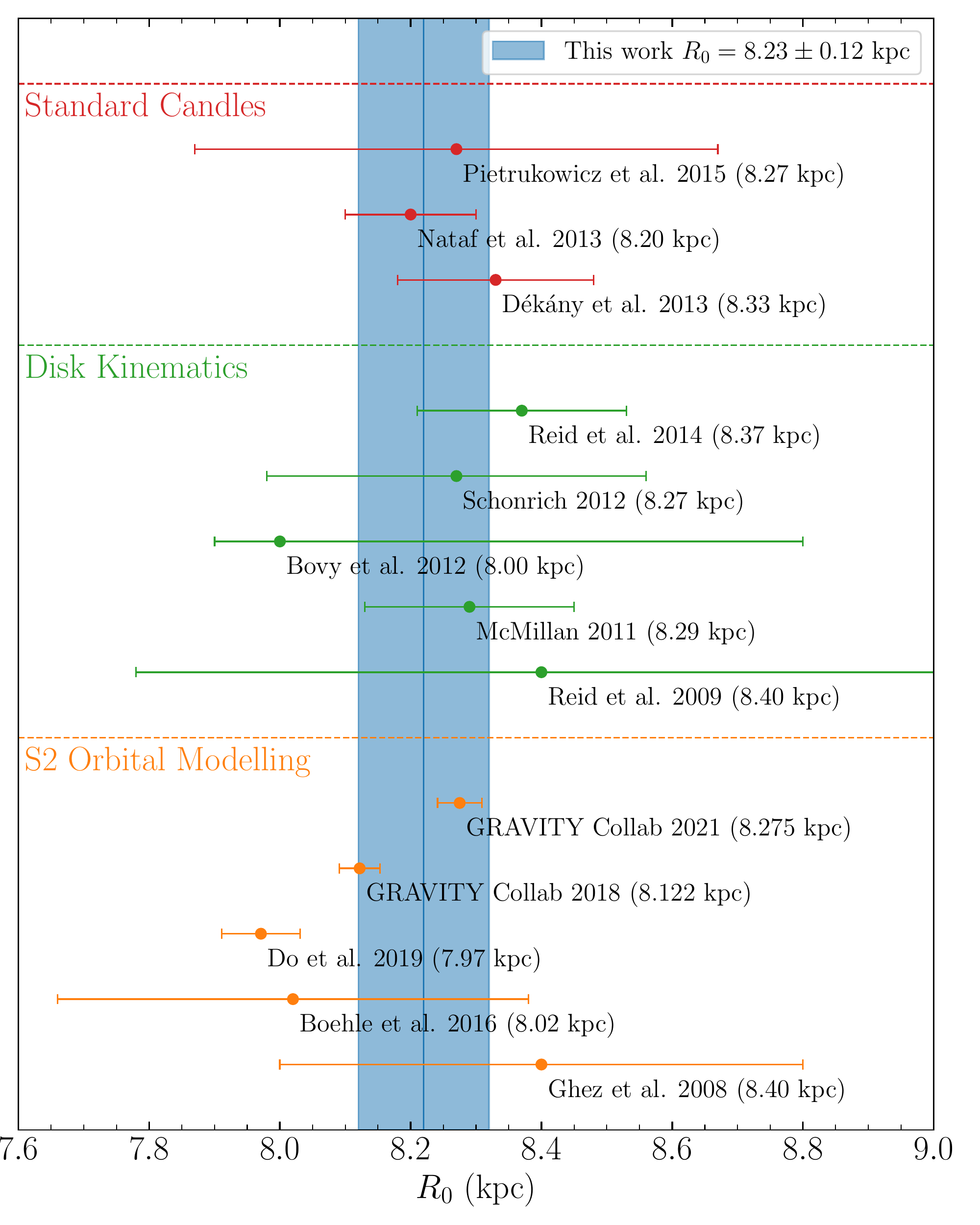}
\caption{A selection of historical $R_0$ measurements using different methods. The vertical blue line and blue colored shaded area show the result and uncertainty from this work.}
\begingroup
    \fontsize{0.01pt}{0.01pt}\selectfont
    \cite{2015ApJ...811..113P} 
    \cite{2013ApJ...776L..19D} \cite{2013ApJ...769...88N} \cite{2009ApJ...700..137R} \cite{2011MNRAS.414.2446M} \cite{2012ApJ...759..131B} \cite{2012MNRAS.427..274S} \cite{2014ApJ...783..130R} \cite{2008ApJ...689.1044G} \cite{2016ApJ...830...17B} \cite{2018A&A...615L..15G} \cite{2019Sci...365..664D} \cite{2021A&A...647A..59G}
\endgroup
\label{fig:r0}
\end{figure}

Because of advances in observational techniques and in dynamical modelling, $R_0$ can now be determined using a variety of methods across the electromagnetic spectrum from radio to X-ray (e.g. \citealt{1984PASJ...36..551E} with distribution of X-ray bursts from compact objects). The resulting $R_0$ measurements range from $8\,\kpc$ to $8.5\,\kpc$ in recent years. A summary of some of the major historical $R_0$ measurements is given in \figurename~\ref{fig:r0}. A catalog of historical measurements of $R_0$ up to the year 2016 and analyses of the trends in the measurements can be found in \citet{2016ApJS..227....5D}. We provide only a brief overview of some of the most recent measurements here. \citet{2011MNRAS.414.2446M} used global dynamical modelling of the Milky Way using multiple observational constraints and obtained $R_0 = 8.29\pm0.16\,\kpc$. Fitting axisymmetric rotation-curve models to radio parallaxes and proper motions of water masers tracing high-mass star-forming regions gave $R_0 = 8.34\pm0.16\,\kpc$ \citep{2014ApJ...783..130R}. A direct radio parallax is not possible for Sgr A$^*$ at present, but is possible for the source Sgr B2 that is near Sgr A$^*$; this gives $R_0 = 7.9\pm 0.75\,\kpc$ \citep{2009ApJ...705.1548R}.

The most precise (but not necessarily most \emph{accurate}) measurements of $R_0$ to date use orbit modelling of the star S2 orbiting the Milky Way's supermassive black hole Sgr A$^*$, most recently giving $R_0 = 7.97\pm0.059\,\kpc$ (from the UCLA group; \citealt{2019Sci...365..664D}) and $R_0 = 8.275\pm0.034 \,\kpc$ (from the MPE group using GRAVITY; \citealt{2021A&A...647A..59G}). The uncertainties in these S2-based measurements are hard to match with other methods, but the significant difference between them demonstrates that systematic uncertainties are still likely larger than estimated by both groups, because issues like aberrations introduced by small optical imperfections along the path from the telescope to the detector in the GRAVITY instrument can have a large effect on the resulting value of $R_0$ (i.e., from $R_0 = 8.122\,\kpc$ in \citealt{2018A&A...615L..15G} to $R_0 = 8.275\pm0.034 \,\kpc$ in \citealt{2021A&A...647A..59G}). More fundamentally, all S2-based (or, more generally, central-parsec-from-Sgr A$^*$-based) $R_0$ determinations have to make the plausible, yet not watertight, assumption that Sgr A$^*$ resides at the barycenter of the Galaxy. Indeed, observational constraints on the offsets between supermassive black holes and the centers of their host galaxies find that $\approx 100\,\pc$ offsets are common (see Fig. 1 and references in \citealt{2021PhRvD.103b3523B})

In this paper, we present a novel approach to determining $R_0$ that manifestly and directly pinpoints the barycenter of the bar+disc system using the kinematics of stars in the central bar region of the Milky Way. Given that the disc dominates the mass distribution out to the solar circle and significantly contributes within the central tens of kpc, this is likely the barycenter of the Milky Way. To do this, we combine data from multiple sources to map the bar's kinematics. APOGEE \citep{2017AJ....154...94M} is an all-sky, high resolution, high signal-to-noise ratio, near-infrared spectroscopic survey within SDSS-IV \citep{2017AJ....154...28B} that combined with the all-sky \gaia\ data allows for detailed studies of the chemo-kinematics of stars in the disc, bar/bulge, and halo components of the Milky Way. Previously, \citet{2019MNRAS.490.4740B} combined  APOGEE DR16 data with \gaia\ DR2 data and obtained precise spectro-photometric distances from \citet{2019MNRAS.489.2079L} generated with a Bayesian neural network. The resulting six-dimensional phase-space data provided an unprecedented view on the Galactic bar region. Now APOGEE DR17 and \gaia\ EDR3 have even greater coverage of and more precise astrometry in the bar region, allowing us to directly map the kinematics of stars near the Galactic center region and determine $R_0$ from it. 

This paper is organized as follows. In \secname\ \ref{sec:data}, we introduce the basic data underlying the analysis and the different samples that we use. \secname~\ref{sec:distances} contains a detailed investigation of systematics in the spectro-photometric distances that we use. We discuss the methodology that we use to determine $R_0$ using our data in \secname~\ref{sec:methodology} and present its results in \secname~\ref{sec:result}. \secname~\ref{sec:discussions} contains a discussion of our results, comparing our $R_0$ measurement to other recent measurements, discussing some implications of our measurement, and we also perform an updated measurement of the bar's pattern speed using our kinematic maps. We summarize our results in \secname~\ref{sec:conclusion} and look forward to the future.

\section{Data}\label{sec:data}

\begin{figure*}
\centering
\includegraphics[width=0.95\textwidth]{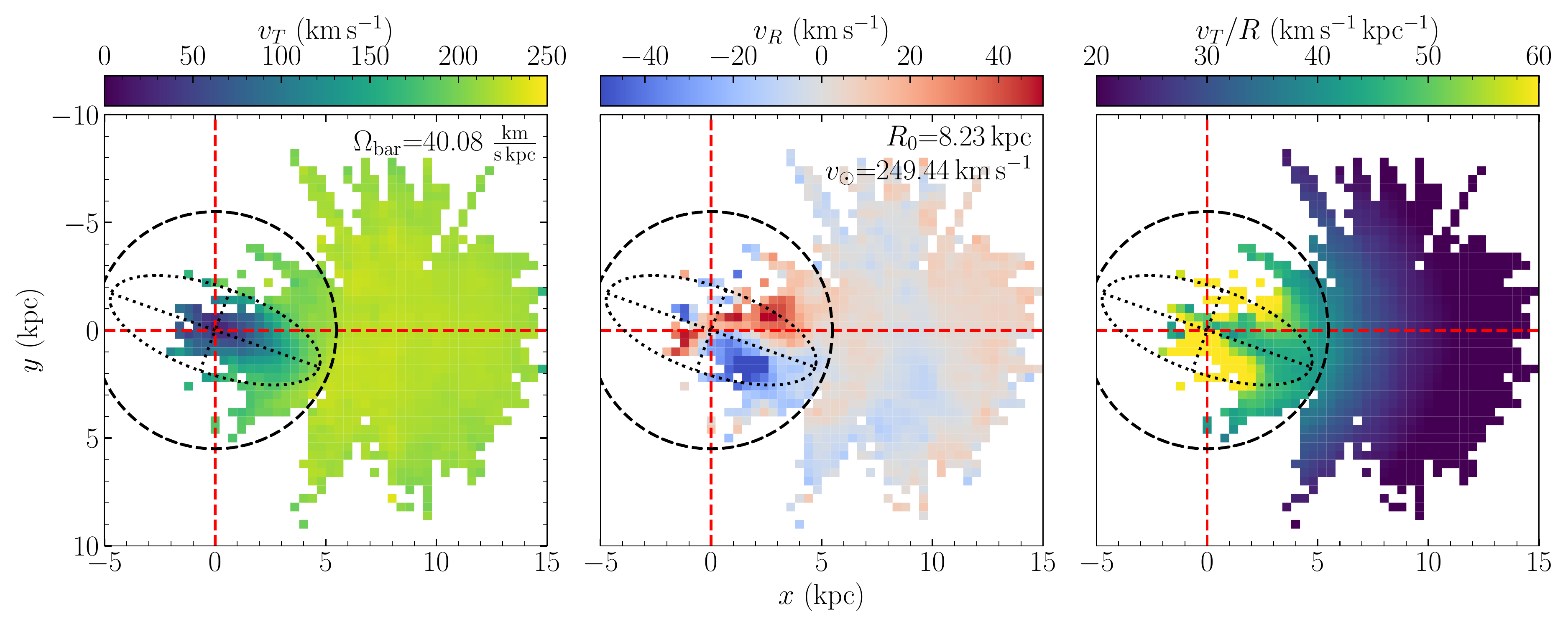}
\caption{Kinematic maps of the bar and disc of the Milky Way using $R_0=\thisr\ \kpc$ and $v_\odot/R_0=30.32\,\kmskpc$. In each panel, the ellipse with its semi-major and semi-minor axis represented by dotted lines shows the bar with a bar angle of 20$^\circ$. The measured pattern speed from these maps is $40.08\pm0.65\,\kmskpc$ (see \secname~\ref{subsec:patternspeed}), which is consistent with \citet{2019MNRAS.490.4740B}. Comparing to the similar plot in \citet{2019MNRAS.490.4740B}, which used APOGEE DR16 and \gaia\ DR2, this plot using APOGEE DR17 and \gaia\ EDR3 shows a clear, bar-shaped global minimum in tangential velocity and clear blue and red regions separated by the semi-major and semi-minor axis in the radial velocity.}
\label{fig:realdata_scenes}
\end{figure*}

To map the kinematics of the bar region with full six-dimensional phase-space, we use a combination of data from \gaia\ EDR3 \citep{2021A&A...649A...1G} and SDSS-IV's APOGEE's DR17 \citep{2017AJ....154...28B,2017AJ....154...94M,2021arXiv211202026A}. Briefly, for a sample of stars selected from APOGEE, we obtain sky coordinates and proper motions from \gaia\ EDR3, line-of-sight velocities from APOGEE, and distances from the \texttt{astroNN} spectro-photometric-distance method from \citet{2019MNRAS.489.2079L} applied to the APOGEE spectra. In this section, we briefly discuss how we use the different data sets to derive the six-dimensional phase-space data that is the basis of our kinematic maps.

The APO Galactic Evolution Experiment (APOGEE) is a near-infrared ($1.5\ \mu m$ to $1.7\ \mu m$), high resolution ($R\sim 22,500$), high signal-to-noise ratio (typical $\mathrm{SNR}>100$ per pixel for $H<12.2$ with 3 hours of integration) large-scale all-sky spectroscopic survey of Sloan Digital Sky Survey (SDSS) mainly targeting red-giants (\citet{2013AJ....146...81Z}; \citet{2017AJ....154..198Z}). The two 300-fiber cryogenic spectrographs \citep{2019PASP..131e5001W} operate at the 2.5-meter Sloan Foundation Telescope \citep{2006AJ....131.2332G} at Apache Point Observatory (APO) responsible for the northern celestial hemisphere and 2.5-meter Irénée du Pont Telescope \citep{Bowen:73} at Las Campanas Observatory (LCO) responsible for southern the celestial hemisphere. We use data from APOGEE's latest data release DR17 \citep{2021arXiv211202026A} where radial velocity, stellar parameters and abundances are derived with the ASPCAP pipeline \citep{2015AJ....150..173N}.

We employ the APOGEE data here for two main purposes: (a) to train, test, and evaluate the \texttt{astroNN} spectro-photometric distances using different subsamples of the data and (b) to obtain the line-of-sight velocity of stars in our kinematic sample. For all APOGEE data that we use, we require a line-of-sight velocity scatter $v_\mathrm{scatter}<1\ \kms$, that no $\texttt{STARFLAG}$ flags are set, and that there is a $K_s$ band apparent magnitude with corresponding extinction available. We perform additional cuts based on the use case that is detailed below. We solely rely on \texttt{astroNN}-determined stellar parameters and abundances \citep{2019MNRAS.483.3255L} for any cuts based on them below.

We use sky coordinates, parallaxes, and proper motions from \gaia\ EDR3. Parallaxes are only used to train the \texttt{astroNN} spectro-photometric distances, but they are not directly used in the production of the kinematic maps of the bar. Sky coordinates and proper motions are only used to create the kinematic maps. For the kinematic APOGEE sample, we obtain \gaia\ EDR3 sky coordinates and proper motions by matching the sky positions with a $2''$ matching radius and then use the closest match (this is unambiguous in all cases); the only additional quality cuts on the \gaia\ EDR3 data for this sample that we apply is that the proper motions cannot be \texttt{Nan}. The proper motion uncertainty is $\approx 0.02\,\masyr$ for $G<15$ at which the bulk of our sample lies. For the parallaxes used in the \texttt{astroNN} training, we perform additional quality cuts and corrections.

To obtain distances for our kinematic sample, we have re-trained our neural-network spectro-photometric distance method \texttt{astroNN} by applying the methods from \citet{2019MNRAS.489.2079L} to the APOGEE DR17 and \gaia\ EDR3 data (\citealt{2019MNRAS.489.2079L} used APOGEE DR14 and \gaia\ DR2). The input data are improved compared with the original training data in the following ways: (i) the number of APOGEE spectra available has increased by almost $300\%$ to 101,435, (ii) the \gaia\ EDR3 parallax uncertainty is smaller by about a factor of two, and (iii) the \gaia\ parallax zero-point is smaller by more than a factor of two as well \citep{2021A&A...649A...4L}. By retraining \texttt{astroNN}, we are able to take advantage of these improvements. The resulting improved distances are included as part of the DR17 \texttt{astroNN} VAC \citep{2021arXiv211202026A}.

To construct the training set, we select APOGEE stars with spectral signal-to-noise $\mathrm{SNR}>200$ in addition to the other cuts discussed above. For the training parallaxes from \gaia\ EDR3, we cut on the Re-normalized Unit Weight Error $\texttt{ruwe}<1.4$, on $\texttt{ip\_frac\_multi\_peak}\leq 2$ and  $\texttt{ipd\_gof\_harmonic\_amplitude}<0.1$ (these cuts indicate that a source is non-single, and on parallax uncertainty $\sigma_\plx<0.1\,\mas$. These cuts are adopted from the recommendations in \cite{2021A&A...649A...5F} for astrometry from \gaia\ EDR3. We apply the parallax zero-point correction from \citet{2021A&A...649A...4L} and do not fit for it separately (as we did in \citet{2019MNRAS.489.2079L}; we discuss the \gaia\ zero-point further in \secname\ \ref{subsec:gplx}. The resulting training set has 101,435 out of all 733,901 APOGEE DR17 stars for training and validation. When training, the parallax is converted to a pseudo-luminosity $L_\mathrm{pseudo} = 10^{\frac{1}{5} m_\mathrm{apparent}} = 10^{\frac{1}{5} M_\mathrm{absolute}+2}$ from \citet{2018AJ....156..145A}, where we use the 2MASS $K_s$ apparent magnitude and extinction \citep{2006AJ....131.1163S}. This pseudo-luminosity preserves the Gaussianity of the parallax uncertainty when we train the neural network model.

After training, the resulting neural-network spectro-photometric distance accuracy is $\approx 5\ \%$. This is better than \gaia\ at large distances ($\approx 5\ \kpc$). Because the distance is the most important uncertain ingredient in our data analysis, we validate our distance determinations in more detail in \secname\ \ref{sec:distances}. When using the \texttt{astroNN} distance for distance validation, we cut on NN \logg\ uncertainty $<0.15\ \dex$, on Re-normalized Unit Weight Error $\texttt{ruwe}<1.4$ to ensure \gaia\ astrometric solution quality, within $500\ \pc$ from the mid-plane and \gaia\ proper motion is not \texttt{NaN}. The rationale for these cuts are discussed in \secname\ \ref{sec:distances}.

The kinematic sample then consists of all stars in APOGEE DR17 that satisfy \logg\ uncertainty $<0.15\ \dex$, $0.8<\logg<3.5$ to cut out stars at the edges of parameter space, Re-normalized Unit Weight Error $\texttt{ruwe}<1.4$, and \gaia\ proper motion is not \texttt{NaN}. This sample has $293,960$ stars in total. To study the kinematics of the bar and disc, we only use stars with vertical distances from the mid-plane less than $500\,\pc$. That sample consists of $138,611$ stars, $16,437$ of which are in the bar region with Galactocentric radii less than 5 kpc.  We convert between heliocentric and Galactocentric coordinates using \texttt{astropy} \citep{2013A&A...558A..33A,2018AJ....156..123A}. To perform this transformation, we need the position and velocity of the Sun in the Milky Way. Everywhere, we will fix the Sun's height above the mid-plane to $z_0=20.8\,\pc$ \citep{2019MNRAS.482.1417B} and the Sun's velocity along the radial and vertical direction to $U_\odot=11.1\,\kms$ and $W_\odot=7.25\,\kms$ \citep{2010MNRAS.403.1829S}. The remaining two parameters, $R_0$ and the Sun's velocity $v_\odot$ in the rotational direction are varied in our analysis, although we will typically fix $v_\odot/R_0 = 30.32\,\kmskpc$ following \citet{2019ApJ...885..131R}. 

We use the kinematic sample to make maps of the median Galactocentric rotational velocity $v_T$, Galactocentric radial velocity $v_R$, and rotational frequency $v_T/R$ in the disc and bar region near the mid-plane. For the data, using what we will find to be the best-fit value of $R_0$ below, these are shown in \figurename~\ref{fig:realdata_scenes}.

\section{Distance accuracy}\label{sec:distances}

\begin{figure}
\centering
\includegraphics[width=0.475\textwidth]{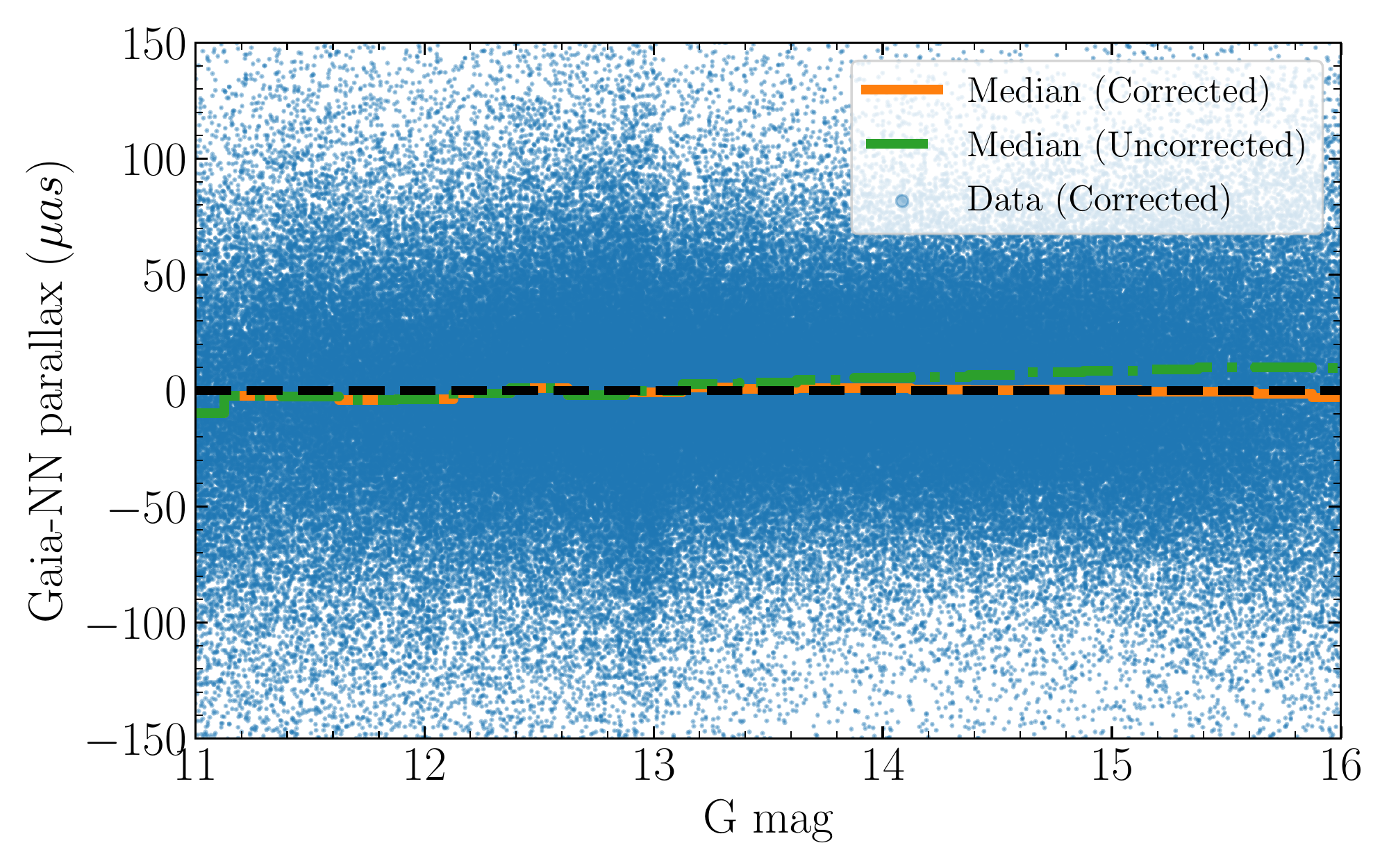}
\caption{Parallax difference between \gaia\ EDR3 and the (inverse) \texttt{astroNN} spectro-photometric distance as a function of \gaia\ $G$ magnitude. The green dotted line shows the median of the parallax difference after adapting the \gaia\ zero-point correction from \citet{2021A&A...649A...4L}, while the orange line includes an additional correction of $5 (G - 13)\uas$ for stars with $13<G<16$ (the black dotted line simply shows $y=0$ for reference). The additional linear correction of the orange line is good for at least $13<G<16$ where stars belong to the same window class in the \gaia\ astrometric calibration pipeline stars with $13 \lesssim
 G \lesssim 16$ belong to WC1).}
\label{fig:zp}
\end{figure}

One of the main factors affecting the accuracy of our $R_0$ measurement is the accuracy of the \texttt{astroNN} neural-network distances that we use. Any systematic bias in the \texttt{astroNN} distances will lead to a similar relative bias in our inferred value of $R_0$. Systematics in the distances can result from deficiencies in the training set or in the neural-network model itself. Because we train using the \gaia\ parallaxes, the \gaia\ parallax zero-point presents a potential issue, especially if the zero-point has complex dependencies on color and ecliptic longitude as shown in \citet{2018A&A...616A..17A}. While in \citet{2019MNRAS.489.2079L}, we also inferred the \gaia\ zero-point as part of training the neural network, we do not do that here because the zero-point correction presented by \citet{2018A&A...616A..17A} turns out to be good enough for our purposes (see below). Other deficiencies in the training set can be an issue, such as incomplete or sparse sampling of important parts of parameter space. For example, the luminous giants that we rely on in APOGEE to investigate the bar region are rare locally and, thus, underrepresented in the training set. This sparseness at the edge of the training set can lead to a neural-network model that performs badly at the edge, as predictions tend to regress to the mean located at the less-luminous giants. We discuss these two possible sources of systematics in the following subsections.

\subsection{\gaia\ parallax systematics}\label{subsec:gplx}

\begin{figure}
\centering
\includegraphics[width=0.475\textwidth]{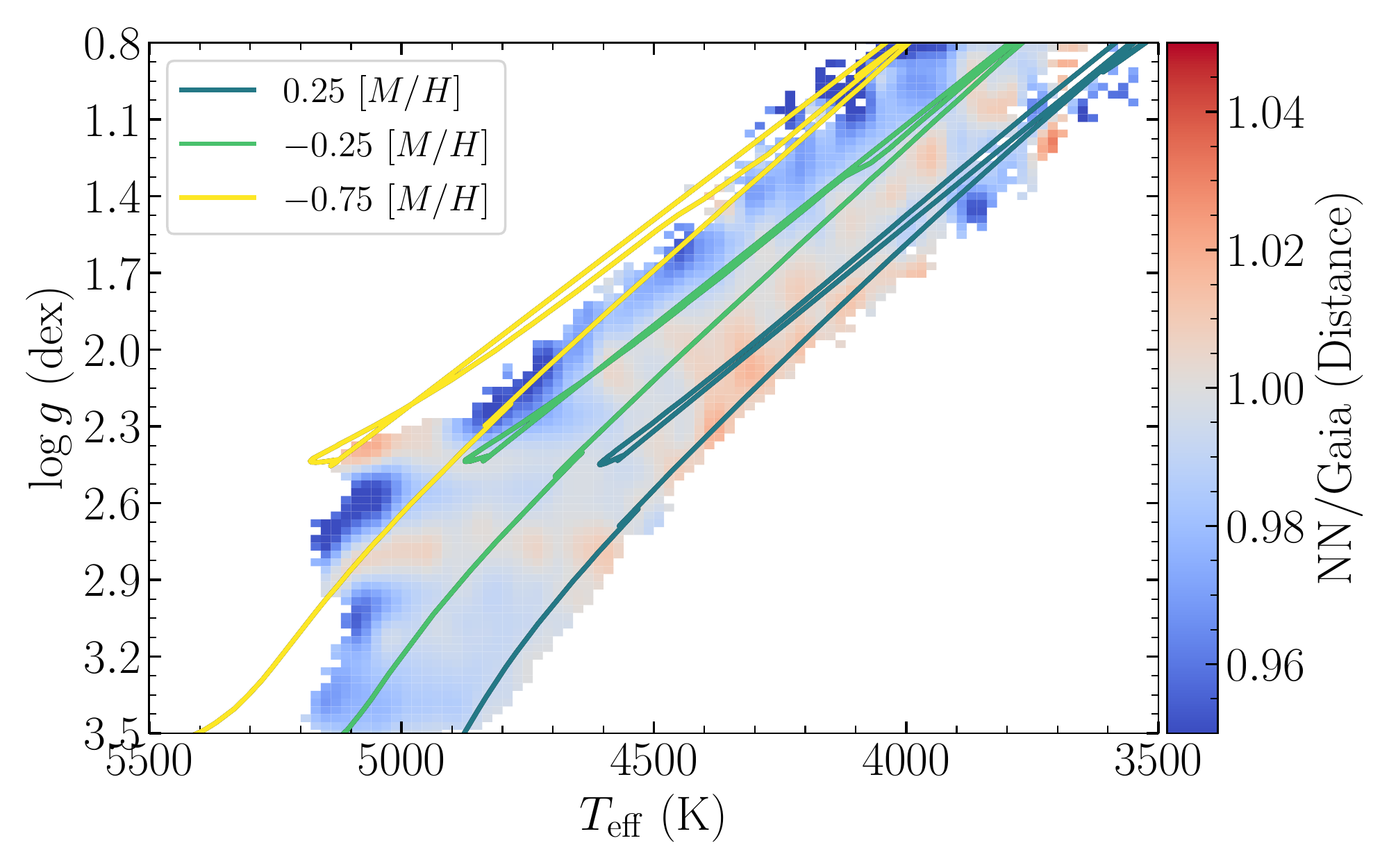}
\caption{Maximum-likelihood ratio from \eqnname~\eqref{eq:ratio_eq} of the \gaia\ parallax to the \texttt{astroNN} parallax as a function of $\teff$ and $\logg$ within $0.5\,\kpc$ of the Galactic plane and including only stars with $G<16$ for reasons discussed in \figurename~\ref{fig:zp} and \secname~\ref{subsec:gplx}. Each bin has a width of $20\,\mathrm{K}$ in $\teff$ and of $0.03\,\dex$ in $\logg$; we only display bins with ten or more stars. PARSEC isochrones \citep{2012MNRAS.427..127B} for an age of $5\ \mathrm{Gyr}$ are over-plotted to give a sense of where $-0.75$ \xh{M} to $0.25$ \xh{M} tracks are located. The \gaia\ parallaxes on average agree with the \texttt{astroNN} parallaxes to about a percent in the well-populated parts of stellar-parameter space and at worst disagree by a few percent.}
\label{fig:teff_logg}
\end{figure}

The \gaia\ parallaxes are known to suffer from a zero-point issue at least partially caused by the variation of the basic angle between the two fields of view that \gaia\ employs \citep{2017A&A...603A..45B} and also because of deficiencies in the data processing pipeline. The zero-point correction is not only important in the training of the \texttt{astroNN} distances, but also in the validation of our distances that we present in this section. Normally, neural networks transfer any systematics present in the training data to the network and, thus, to any outputs derived from it. But we are not directly training on the \gaia\ parallaxes when training the \texttt{astroNN} model, instead the model predicts a pseudo-luminosity that needs to be combined with a star's apparent magnitude to obtain the distance. Because the \gaia\ zero-point cannot be predicted directly from the APOGEE spectra, the \texttt{astroNN} will transfer the average effect of the parallax zero-point as a result. However, this could still lead to systematic biases.

Unlike in \citet{2019MNRAS.489.2079L}, we fix the \gaia\ parallax zero-point during the training rather than fitting for it and we fix it to the model from \citet{2021A&A...649A...4L}. Numerous other works have independently validated the accuracy of the \citet{2021A&A...649A...4L} correction. For example, \citet{2021AJ....161..214Z} validated the correction using stars with astrometric distances in the APOGEE-Kepler field and they report that the \citet{2021A&A...649A...4L} zero-point correction for the five-parameters astrometric solution, which is the relevant case for almost all APOGEE stars, is good within $2\ \uas$ at $G<13$. However, they do not cover $G > 13$ due to a lack of stars at these magnitudes in their sample, which is not ideal because we do depend on these stars and the \gaia\ instrument and pipeline processes data at these magnitudes differently due to the use of astrometric window classes. Stars with $G>13$ are particularly of interest due to the high level of extinction towards the bar. \citet{2021ApJ...910L...5H} validated the \gaia\ zero-point correction using LAMOST primary red-clump stars and they report that the zero-point over-corrects the parallaxes for stars with $G>14$ in the five-parameters astrometric solution.

To investigate the appropriateness of the \citet{2021A&A...649A...4L} correction, \figurename\ \ref{fig:zp} shows the parallax difference between \gaia\ and our \texttt{astroNN} distances as a function of \gaia\ $G$ magnitude. The parallax zero-point is known to be a non-trivial function of magnitude, color, sky position, etc. Most of the zero-point's dependencies cannot be inferred from spectra alone, thus, they do not affect the resulting spectro-photometric distances (discussed extensively in \citealt{2019MNRAS.489.2079L}). Without the presence of any parallax bias there should be no median difference between \gaia\ and \texttt{astroNN} parallaxes, but any parallax bias from \gaia\ will show up as a trend as a function of zero-point dependencies. Thus, by looking for trends in the difference between \gaia\ and \texttt{astroNN} parallaxes, we can test for the presence of remaining zero-point systematics in \gaia. The green line is the median as a function of $G$ and we observe an over-correction at faint end similar to that found by \citet{2021ApJ...910L...5H}. The over-correction is clearly seen to start at $G>13$, which is unsurprising as it is a break-point in \gaia's astrometric analysis. In \figurename\ \ref{fig:zp}, the orange dotted line (median) and the blue scattered points (individual stars) both display the difference after an additional correction of $5 (G - 13)\,\uas$ for stars with $13 < G < 16$. It is clear that this succeeds in flattening the green curve. Thus, we find that the \citet{2021A&A...649A...4L} correction should be slightly altered at $13 < G < 16$. While ideally we would apply this new correction before training the \texttt{astroNN} distance network, stars with $13 < G < 16$ are sufficiently rare in the training set, because of our signal-to-noise ratio cut, that the trained network is hardly affected by this change. We have explored retraining the model after applying the additional correction, but find no significant difference in the predicted distances. Therefore, we continue to use the original distances in this paper (these are the ones in the official \texttt{astroNN} VAC and this choice therefore makes it easier to reproduce our results). However, we do apply the additional correction in the next section when validating our \texttt{astroNN} distances, because stars with $13 < G < 16$ \emph{are} common at large distances and while the \texttt{astroNN} distance does not depend on the additional correction, the \gaia\ parallaxes that we use to validate our distances \emph{do} strongly depend on them.

\begin{figure}
\centering
\includegraphics[width=0.475\textwidth]{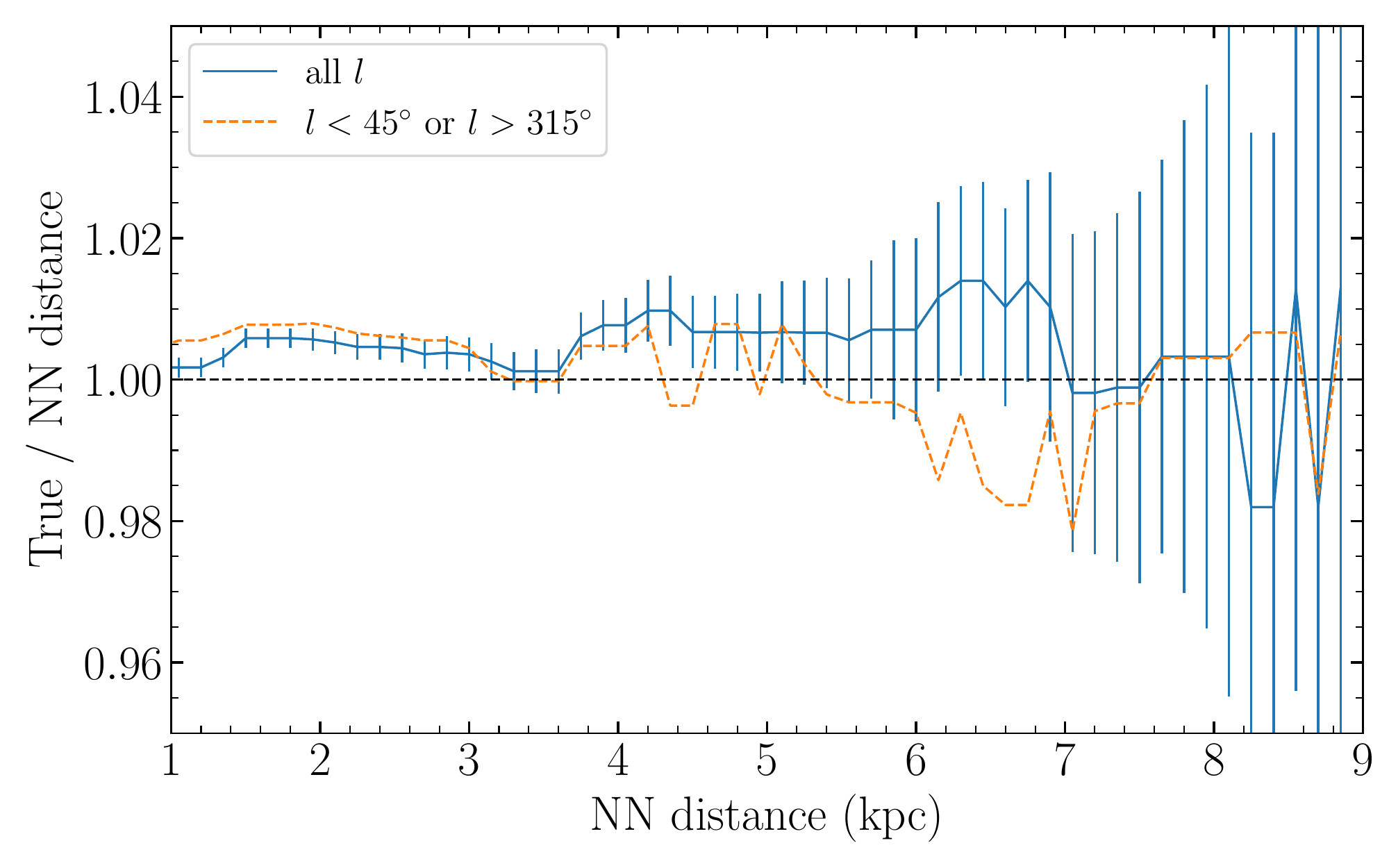}
\caption{Ratio of the true distance to the \texttt{astroNN} spectro-photometric distance as a function of \texttt{astroNN} distance ranging from $1\,\kpc$ to $9\,\kpc$ within $0.5\,\kpc$ of the Galactic mid-plane. Only stars with $G<16$ are used and an additional zero-point correction of $5(G-13)\uas$ for stars with $13<G<16$ is applied ton the \gaia\ parallaxes for the reasons discussed in \figurename\ \ref{fig:zp} and \secname~\ref{subsec:gplx}. The true distance is the distance inferred from a robust maximum-likelihood stacking of the \gaia\ parallaxes in $0.15\,\kpc$ wide bins in \texttt{astroNN} distance (\eqnname~\ref{eq:bias_eq}). The error bars are statistical uncertainty determined by bootstrap re-sampling; they are correlated because neighbouring bins overlap. The blue line employs all of the stars, while the orange line only uses stars within the direction to the bar. The ratio is almost always within $1\%$ for both lines showing that the \texttt{astroNN} distance accuracy is $\lesssim 1\%$.}
\label{fig:systematic}
\end{figure}

\subsection{\texttt{astroNN} distance systematics}\label{subsec:systematic}

\begin{figure*}
\centering
\includegraphics[width=0.95\textwidth]{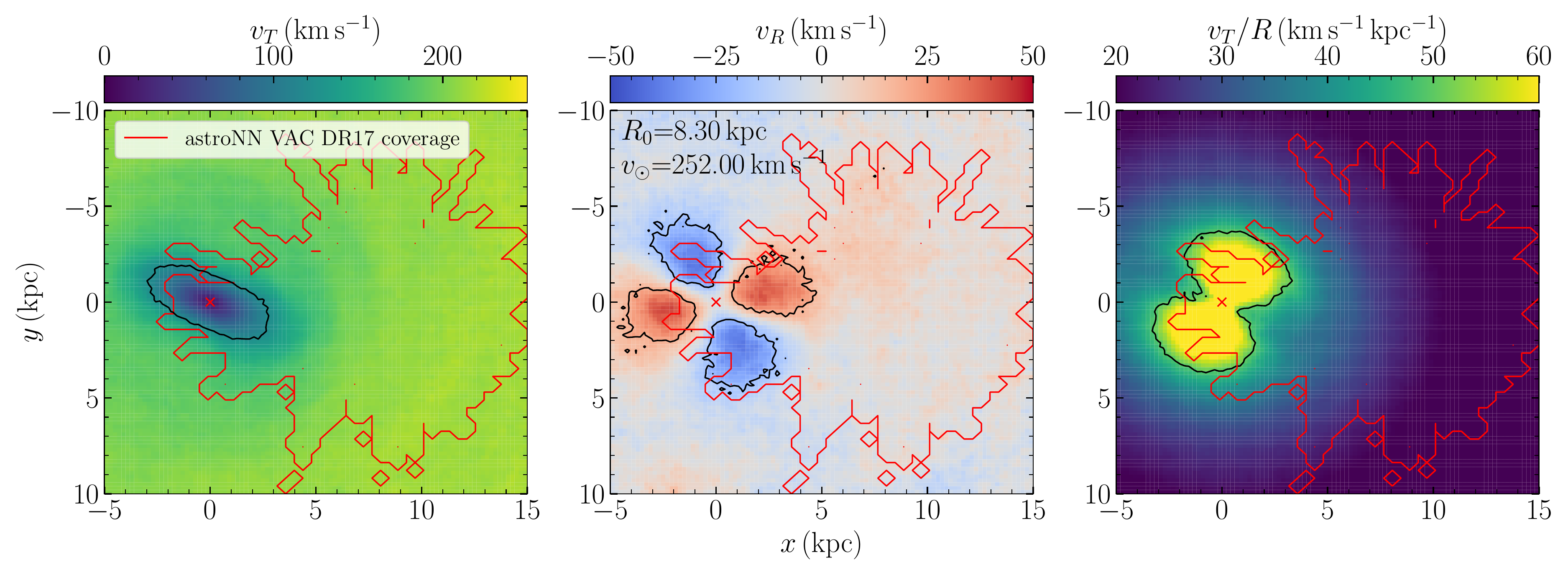}
\caption{Kinematic maps of the bar and disc for the simulation from \citet{2017MNRAS.464..702K}. The bar pattern speed ground truth is $24.5\,\kmskpc$ at a $25^{\circ}$ bar angle. Each bin has a width of $0.2\,\kpc$. The red contours represent the area covered by the APOGEE data in the Milky Way.}
\label{fig:sims}
\end{figure*}

The neural-network model used to train and evaluate the \texttt{astroNN} distances can introduce distance systematics, for example, if important parts of parameter space are sparsely covered by training data or at the edges of parameter space, where predictions will tend to regress to the mean. Because we depend on luminous giants in APOGEE to reach the Galactic bar region, which are rare in the training sample with high signal-to-noise ratio spectra and located near the edge of parameter space, both of these effects are highly relevant.

We investigate the presence of any systematic biases in the \texttt{astroNN} distances in this sub-section. \figurename\ \ref{fig:teff_logg} displays the ratio between the \gaia\ parallax with the additional $G$-mag dependent correction shown in \figurename\ \ref{fig:zp} and the inverse \texttt{astroNN} distance (i.e., the \texttt{astroNN} parallax). We robustly determine the ratio in bins of \teff\ and \logg\ by assuming that stars in the same \teff\-\logg\ bin have similar luminosities. The space of \teff\ and \logg\ is a logical choice of parameter space, because the neural-network model predicts spectro-photometric distances from spectra, and therefore, features like \teff, \logg, and metallicity dominate the features used by the network. The feature \logg\ is a particularly important for predicting the luminosity. In the absence of any distance bias, we expect that the shown parallax ratio is one in each bin. Our robust estimate of the ratio in each bin is obtained using a maximum-likelihood-like method by minimizing 
\begin{equation}\label{eq:ratio_eq}
\min_{\Delta} \qquad \sum_{i}{ \left|\frac{\plx_\mathrm{i, gaia}}{\plx_\mathrm{i, NN}}-\Delta \right| \frac{1}{\sigma_{\plx, i}}}
\end{equation}
for $\Delta$, the inferred parallax ratio. Most stars are within only a few percent in  \figurename~\ref{fig:teff_logg} except in some parts of the metal-poor end of the diagram that is sparsely covered in the training set. The median of $|1-\Delta|\approx1\%$ for the bins displayed in \figurename~\ref{fig:teff_logg}.

To test in more detail for the presence of distance biases in the Galactic bar region, we also examine how distance biases vary as a function of distance in \figurename~\ref{fig:systematic}. In this figure, we present a similar check as first performed by \citet{2019MNRAS.490.4740B}: we assume that if \texttt{astroNN} states that two stars are at the same distance, they truly are so, even if \texttt{astroNN} might return a biased distance. Thus, we can create large samples of stars at a given distance. For these, we can statistically stack the \gaia\ parallaxes (corrected for zero-point offsets) and determine the true, geometric distance that provides a good check for distance biases. Specifically, for stars in a bin of \texttt{astroNN} distance, we infer the \gaia\ distance by minimizing 
\begin{equation}\label{eq:bias_eq}
\min_{D} \qquad \sum_{i} \frac{|\plx_{i} - 1/D|}{\sigma_{\plx, i}}
\end{equation}
where $D$ is the true distance that minimize the equation for each distance bin. \figurename~\ref{fig:systematic} displays the resulting ratio between the true and the \texttt{astroNN} distance as a function of \texttt{astroNN} distance for all stars within $0.5\,\kpc$ from the mid-plane as well as for those stars towards the bar region. The ratio is almost always within $1\%$ in both cases.

The tests shown in \figurename~\ref{fig:teff_logg} and \figurename~\ref{fig:systematic} demonstrate that our \texttt{astroNN} distances have systematic biases $\lesssim 1\%$, or about $80\,\mathrm{pc}$ at the distance of the Galactic center.

\section{Methodology}\label{sec:methodology}

Stars in the Milky Way disc and bar rotate around the center with a mean tangential velocity $\langle v_T \rangle$. The $m=2$ symmetry of the bar requires that $\langle v_T \rangle$ reaches an extremum at the location of the barycenter of the Milky Way. Simulations such as the ones that we discuss below demonstrate that this extremum is a minimum at $\langle v_T \rangle \approx 0$ for galaxies with bars similar to that of the Milky Way. It is this minimum in $\langle v_T \rangle$ that we use to determine $R_0$.

\figurename~\ref{fig:sims} shows similar kinematic maps of the mean $v_T$, $v_R$, and $v_T/R$ for the simulation of a barred Milky-Way-like galaxy from \citet{2017MNRAS.464..702K}. This simulation does not include a supermassive black hole at the center or gas, but a comparison with the kinematic maps of the data in \figurename~\ref{fig:realdata_scenes} demonstrates that they are similar and the simulation therefore matches the dynamics in the bar region well despite its limitations. We have oriented the bar at an angle of $25^\circ$ to approximately match the bar angle in the Milky Way. \figurename~\ref{fig:sims} displays the data correctly, that is, without distorting them by using an incorrect value of $R_0$. We also overlay the approximate footprint of the data kinematic maps (which is driven by APOGEE's selection function). We see multiple signatures pinpointing the location of the galactic center. The mean rotational velocity $v_T$ shows a well-defined minimum at the center, while the radial velocity $v_R$ displays a saddle point, and the rotational frequency $v_T/R$ a maximum. The radial velocity map also displays a distinct clover-leaf pattern of high and low radial velocities within the bar region. This clover-leaf pattern also locates the galactic center. However, it does not do so as clearly as the minimum in $v_T$.

\begin{figure*}
\centering
\includegraphics[width=0.95\textwidth]{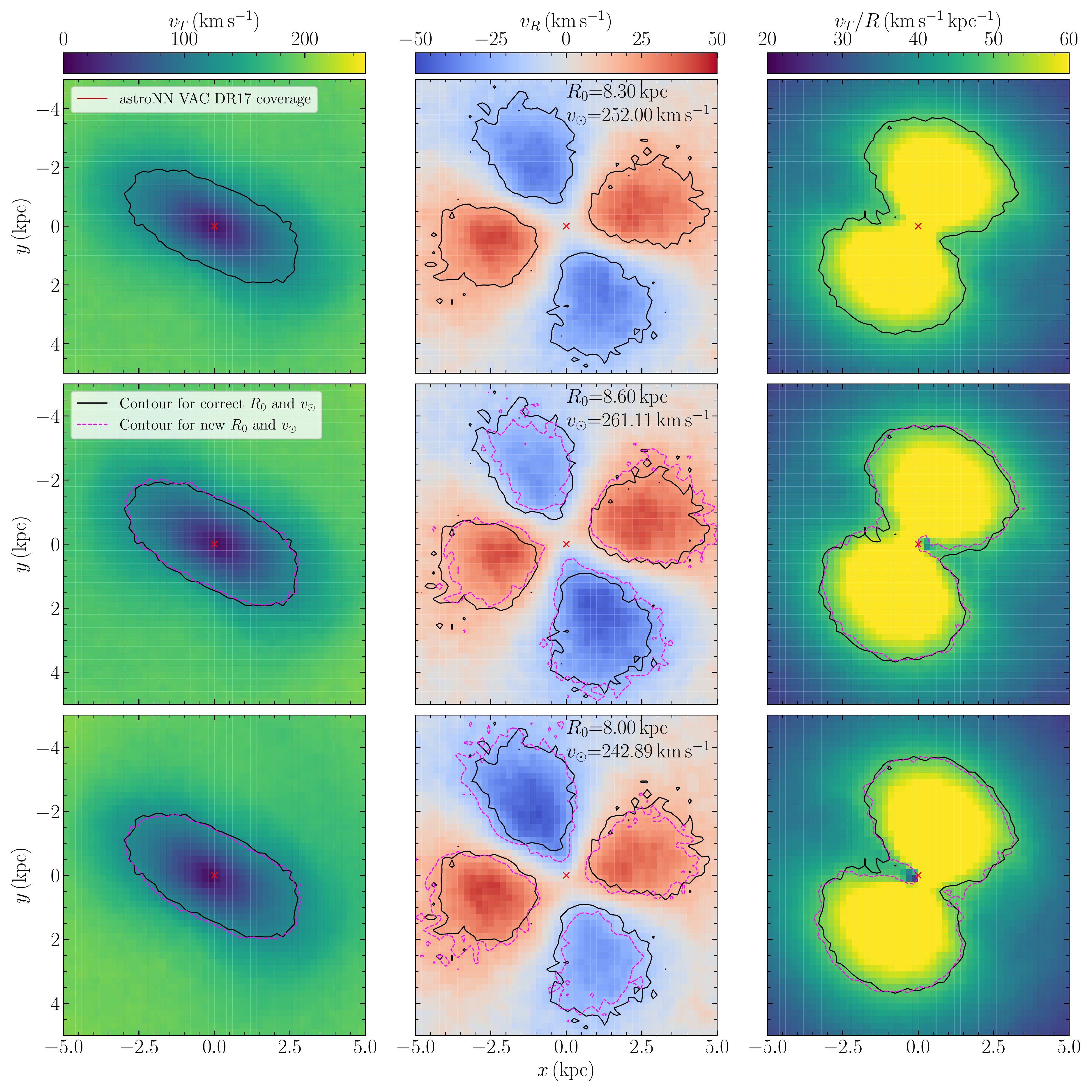}
\caption{Kinematic maps like in \figurename~\ref{fig:sims}, but zooming in on the central region and including the effects of assuming the wrong galactic parameters. The different rows use different sets of $R_0$ and $v_\odot$ while keeping $v_\odot / R_0$ fixed at $\approx 30.32\,\kmskpc$. The top row uses the true $R_0$ and $v_\odot$ ($R_0=8.3\,\kpc$ and $v_\odot252.00\,\kms$). The second row uses an $R_0$ offset by $0.3\,\kpc$ and $v_\odot=261.11\,\kms$, while the third row uses an $R_0$ offset by $-0.3\,\kpc$ and $v_\odot=242.89\,\kms$. The red cross in each panel always highlights the location of the center at $(0, 0)$. The black contours in all three rows shows the contours of $120\,\kms$ for $v_T$, $20\,\kms$ for $v_R$, and $50\,\kmskpc$ for $v_T/R$ using the correct set of $R_0$ and $v_\odot$, while in the second and third row, the purple contours represent the same levels for the incorrect set of $R_0$ and $v_\odot$.}
\label{fig:sims_2}
\end{figure*}

\begin{figure*}
\centering
\includegraphics[width=0.95\textwidth]{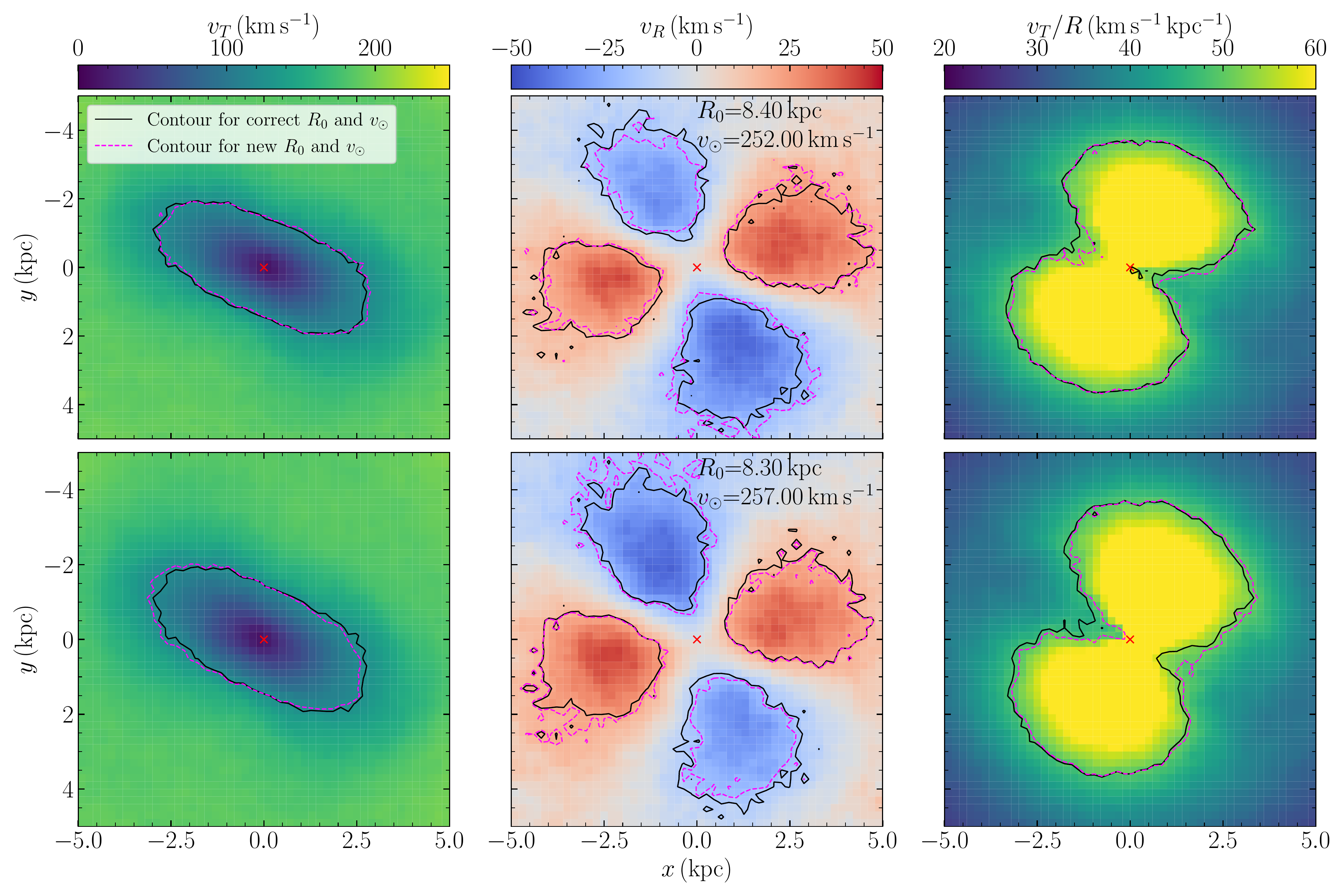}
\caption{Like the middle and bottom row of \figurename~\ref{fig:sims_2}, but changing $R_0$ and $v_\odot$ without keeping $v_\odot/R_0$ constant. Comparing to \figurename~\ref{fig:sims}, a small offset in either $R_0$ or $v_\odot$ produces more significant shifts in the  kinematic maps than offsets in $R_0$ and $v_\odot$ that keep $v_\odot / R_0$ fixed.}
\label{fig:sims_3}
\end{figure*}

\figurename~\ref{fig:sims_2} contains a zoomed-in version of \figurename~\ref{fig:sims} as the top row and the middle and bottom row shows what happens when we create the kinematic maps of the simulation using wrong values of $R_0$ and the Sun's rotational velocity $v_\odot$---this $v_\odot$ is the second most important parameter that has to be assumed to make the kinematic maps in the Galactocentric frame---but keeping the value of $v_\odot/R_0$ constant. That is, we convert the simulation's phase-space data to the heliocentric coordinates of a mock observer located at $R=8.3\,\mathrm{kpc}$, $\phi=z=0$, $v_\odot = 252\,\kms$, and $U_\odot=W_\odot=0$. We then convert these coordinates back to Galactocentric coordinates assuming a wrong value of $R_0$ and $v_\odot$.

The left panels clearly demonstrate that the location of the minimum shifts away from $(0,0)$, while the middle panels show that the saddle point in $v_R$ remains approximately at $(0,0)$. The change in the $v_R$ pattern is therefore more subtle and, thus, more difficult to detect: the leafs of the clover become asymmetric (compare the purple contours of simulations with the wrong parameters to the symmetric contours of the simulation with the correct parameters), but this is difficult to detect without full coverage of the bar region. While the maximum in $v_T/R$ does shift away from $(0,0)$, we do not use it, because (i) it does not contain different information from $v_T$ and (ii) the maximum's value is less universal.

\figurename~\ref{fig:sims_3} shows another two sets of wrong values for $R_0$ or $v_\odot$, but this time not keeping the ratio $v_\odot/R_0$ constant. Comparing to \figurename~\ref{fig:sims}, we see that the changes to the contours in $v_R$ and $v_T/R$ are similar, even though the changes to $R_0$ and $v_\odot$ are much smaller than those in \figurename~\ref{fig:sims}. Thus, when $v_\odot/R_0$ is well constrained, constraining $R_0$ using $v_R$ or $v_T/R$ is more difficult, while the minimum in $v_T$ remains sensitive to $R_0$ shifts.

\begin{figure}
\centering
\includegraphics[width=0.475\textwidth]{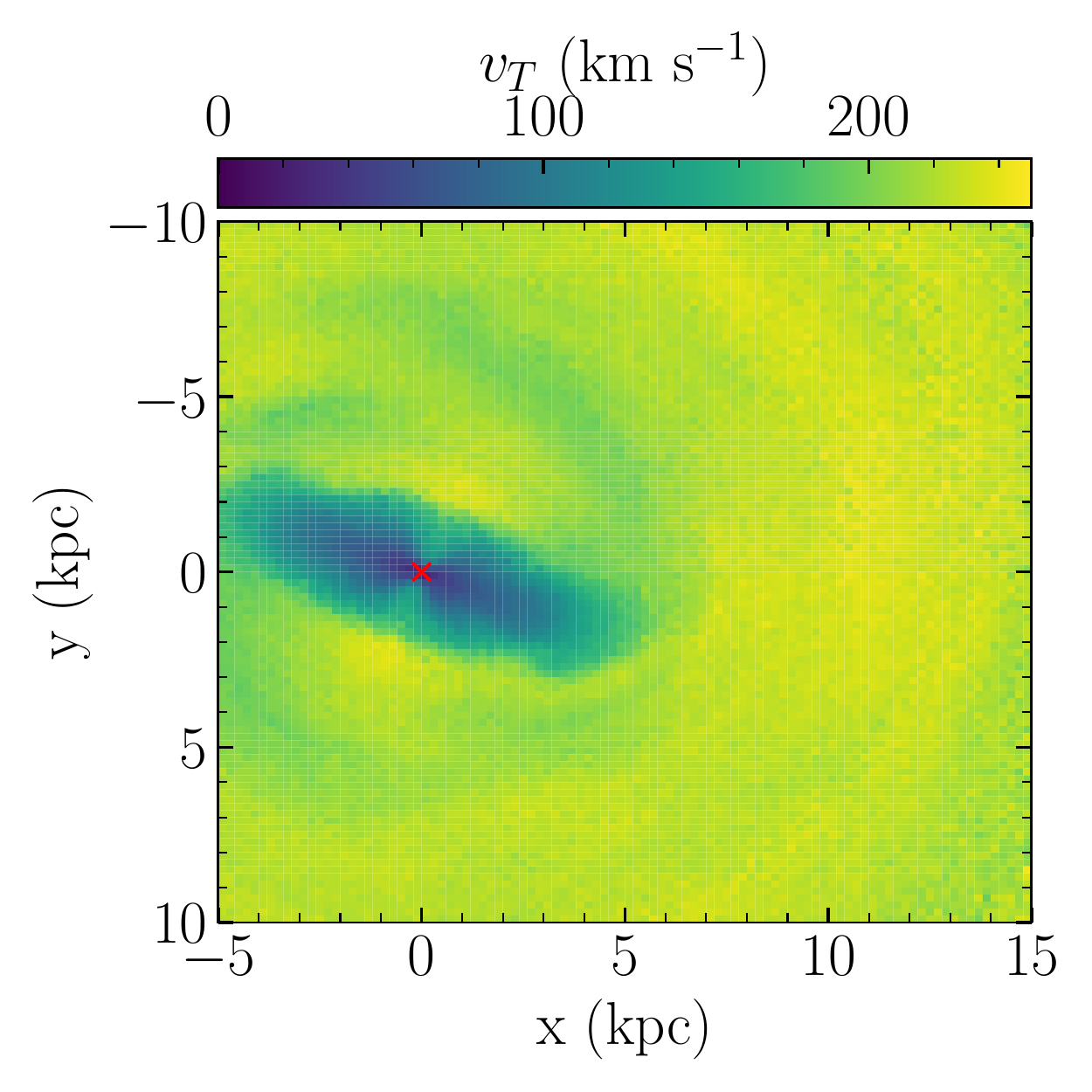}
\caption{Kinematic $v_T$ map from a Milky-Way-like simulation Milky-Way disc presented in \citet{2021arXiv210708055B}. The simulated disc is perturbed by the infall of a massive Sgr-like dwarf. The red cross show the location of the galactic center at $(0, 0)$. The figure shows that $v_T$ is approximately constant near the center without showing as clear of a minimum as for the simulation presented in \figurename~\ref{fig:sims} or as in the real data from \figurename~\ref{fig:realdata_scenes}. Nevertheless, applying our fitting routine returns the correct value of $R_0$ for this simulation, albeit with larger uncertainties than for the simulation in \figurename~\ref{fig:sims}.}
\label{fig:morgansims}
\end{figure}

To investigate the universality of the minimum in $v_T$ that we observe in \figurename~\ref{fig:sims}, we also consider kinematic maps created from the simulation suite presented by \citet{2021arXiv210708055B}, which includes unperturbed Milky-Way-like galaxies as well as galaxies perturbed by the Sgr dwarf galaxy. \figurename~\ref{fig:morgansims} shows an example kinematic $v_T$ map from one of their simulations, which is perturbed by the infall of the massive Sgr dwarf. It is clear from this figure that this simulated bar does not have as clear of a minimum as the simulation in \figurename~\ref{fig:sims} and there is some structure along the minor axis that is not present in the simulations from \citet{2017MNRAS.464..702K}. Nevertheless, running the fitting routine described below on the data in \figurename~\ref{fig:morgansims} returns the correct value of $R_0$ albeit with larger uncertainties. The other simulations in the suite from \citet{2021arXiv210708055B} give similar results. Because the observed kinematic maps resemble those from \figurename~\ref{fig:sims} more than they do that in \figurename~\ref{fig:morgansims} (which is also seen by \citealt{2021arXiv211203295E} using the spectro-photometric distances from \citealt{2019AJ....158..147H}), they provide a more realistic test of our method.

The current footprint of the data (shown by the red contours in \figurename~\ref{fig:sims}) clearly limits our use of the data: as discussed above, detecting the subtle changes in the clover-leaf pattern in $v_R$ is difficult with this footprint, and even in $v_T$ we do not have a wide range of Galactocentric azimuths. To keep things simple, therefore, we only use $v_T$ along the $y=0$ axis, that is, we use a one-dimensional slice of the kinematic maps along the center-Sun line. The main advantage of this is that it makes the creation of an objective function to minimize simple.

To find the minimum in $v_T$, we calculate the smoothed running median using sliding window of $v_T$ along the x-axis (with $|y| < 0.5\,\mathrm{kpc}$) for the stars in our kinematic sample within $500\ \pc$ from mid-plane. Then we find the value of $R_0$ that minimizes the following objective function:
\begin{equation}\label{eq:objective}
\min_{R_0} \qquad f\left(R_0, v_\odot[R_0]\right) = \sqrt{\left(\frac{x_i}{2\ \kpc}\right)^2+\left(\frac{v_{T, i}}{50\ \kms}\right)^2}
\end{equation}
where the scaling parameters $2\ \kpc$ and $50\ \kms$ are chosen to approximately scale both $R$ and $v_T$ onto the same scale. The rotational velocity $v_{T, i}$ is the value of the running median at positions $x_i$ indexed by $i$. We always use $R_0$ and $v_\odot/R_0$ as the independent parameters, so $v_\odot(R_0) = R_0 \times v_\odot/R_0$ when we vary the assumed value of $v_\odot/R_0$. 

\begin{figure}
\centering
\includegraphics[width=0.475\textwidth]{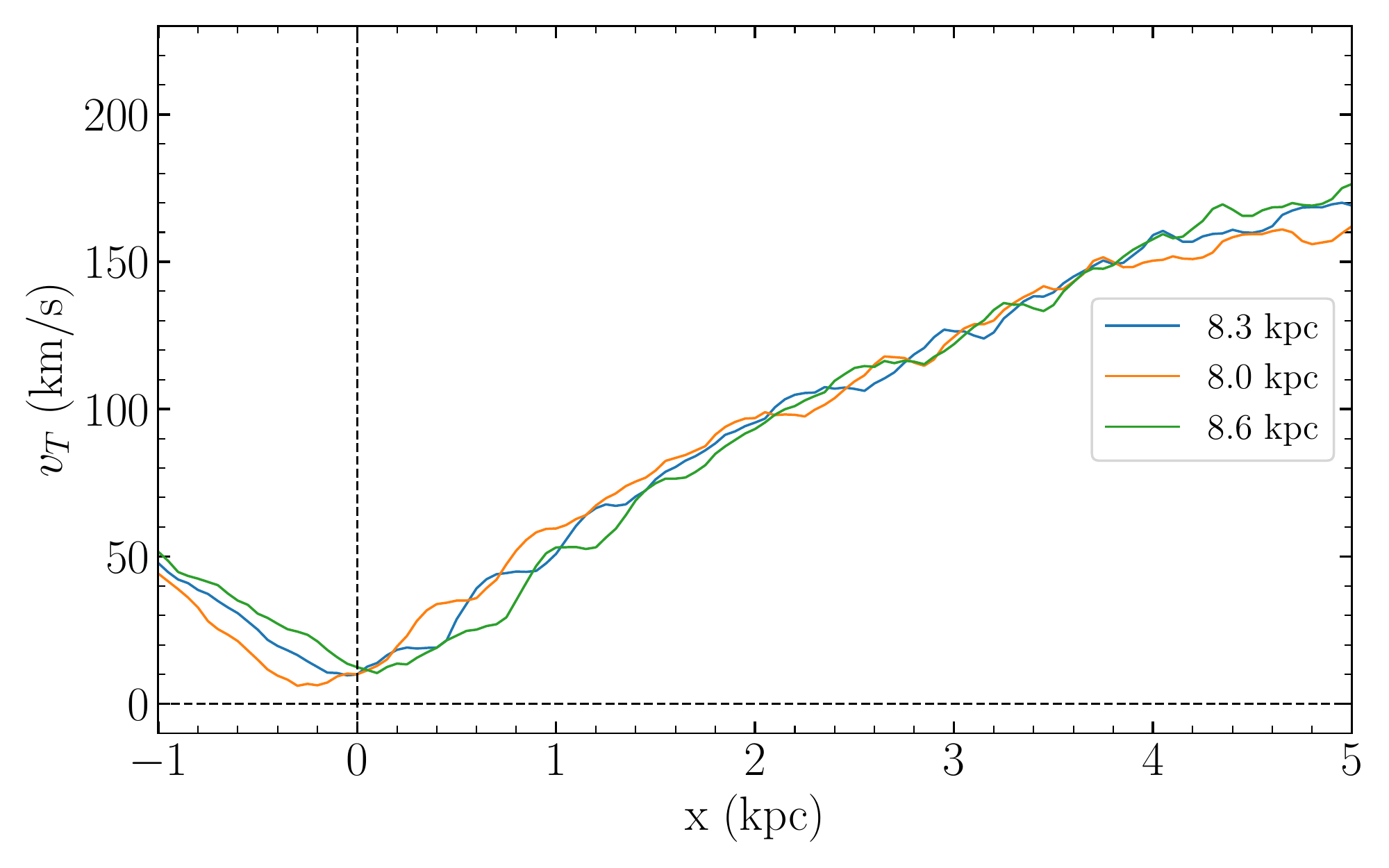}
\caption{Median tangential velocity $v_T$ for stars in the simulation from \citet{2017MNRAS.464..702K} that are $\pm 0.5\,\kpc$ from the $x$-axis smoothed by a Gaussian kernel. The median velocity $v_T$ is calculated for $R_0 = 8.0\,\kpc$, $8.3\,\kpc$ (the correct value), and $8.6\,\kpc$. For the correct $R_0$, a minimum in $v_T$ is observed near the origin. For incorrect values of $R_0$, the minimum occurs away from $x=0$ and at larger values of $v_T$. This figure is similar to \figurename~\ref{fig:vT_original} for the real data.}
\label{fig:vT_sims_original}
\end{figure}

We apply this fitting routine to the simulations from \citet{2017MNRAS.464..702K} shown in \figurename~\ref{fig:sims}, using different true values of $R_0$ for the heliocentric observer in the simulation. The value of the median $v_T$ as a function of $x$ for three different true values of $R_0$ is shown in \figurename~\ref{fig:vT_sims_original}, where the correct minimum should occur at $x=0$. This figure demonstrates that we successfully recover $R_0$ to an accuracy of $\approx 0.05\,\kpc$.

\section{Result}\label{sec:result}

We determine $R_0$ using the kinematic sample by optimizing the objective function in \eqnname~\eqref{eq:objective} and we obtain its random uncertainty by bootstrapping the kinematic sample (that is, we create new kinematic samples of the same size by re-sampling with replacement from the original one). To account for the uncertainty in the angular Galactocentric velocity $v_\odot/R_0$ of the Sun, we use the measurement of $v_\odot/R_0=30.32\pm0.27\ \kmskpc$ from \citet{2019ApJ...885..131R} in the following way. To determine our best-fit value of $R_0$, we fix $v_\odot/R_0=30.32\,\kmskpc$, but when determining the uncertainty in $R_0$, we sample from a Gaussian distribution with a mean of $v_\odot/R_0=30.32\,\kmskpc$ and a standard deviation that is eight times the uncertainty from \citet{2019ApJ...885..131R}: $\sigma_{v_\odot/R_0} = 2.16\,\kmskpc$. We use eight times the stated uncertainty as the standard deviation to be very conservative in our assumption about $v_\odot/R_0$. Because the measurement from \citet{2019ApJ...885..131R} is based on the kinematics of masing high-mass star-forming regions in the disk, it is independent from Sgr A$^*$ and our adoption of $v_\odot/R_0$ from \citet{2019ApJ...885..131R} therefore does not introduce any dependence on Sgr A$^*$ being at the center of the Milky Way.

\begin{figure}
\centering
\includegraphics[width=0.475\textwidth]{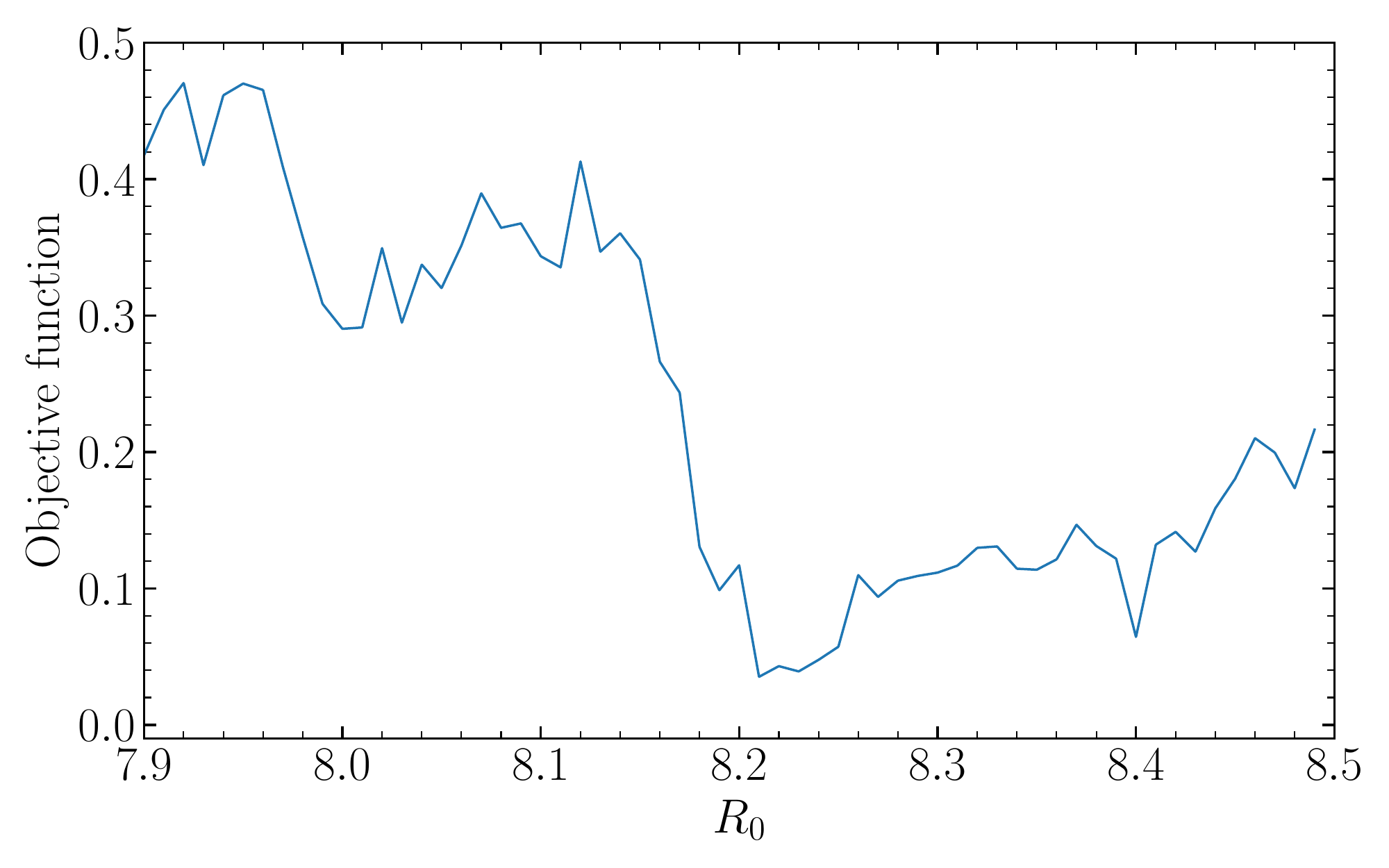}
\caption{Objective function from \eqnname~\ref{eq:objective} for a range of possible values of spanning the range of historical measurements. The objective function displays a minimum at $R_0 =  \thisr\pm\thisrerr\,\kpc$. Values of $R_0 \lesssim 8.15\,\kpc$ are especially disfavored.}
\label{fig:objective}
\end{figure}

The resulting measurement is $R_0=\thisr\pm\thisrerr\,\kpc$. \figurename~\ref{fig:objective} shows the objective function for $v_\odot/R_0=30.32\,\kmskpc$. It displays a clear minimum at $R_0 \approx 8.2\,\kpc$. Values of $R_0 \lesssim 8.15\,\kpc$ in particular lead to large values of the objective function and are, thus, disfavored. This clearly rules out the lower end of historical $R_0$ measurements shown in \figurename\ \ref{fig:r0}. 

\begin{figure}
\centering
\includegraphics[width=0.475\textwidth]{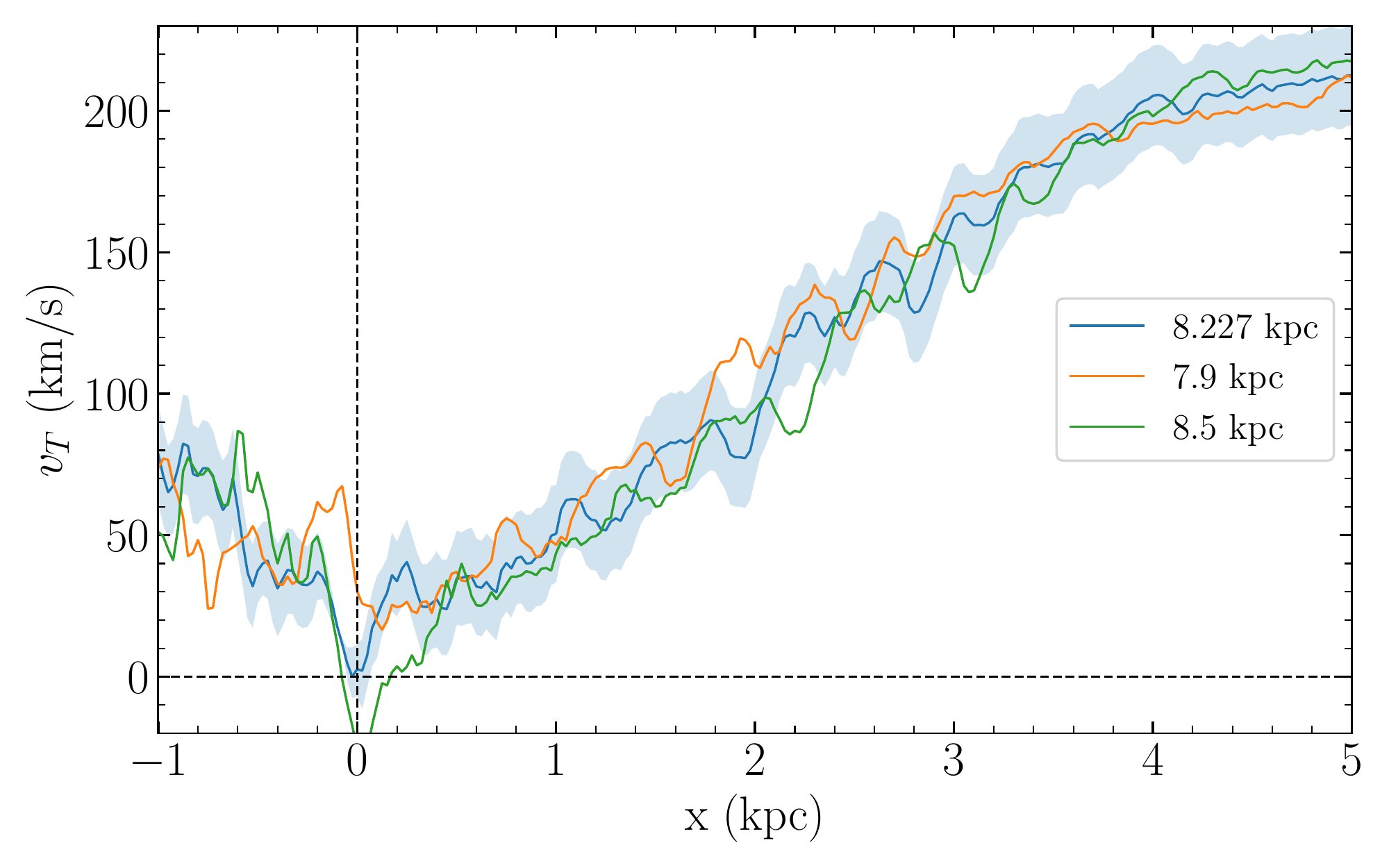}
\caption{Median tangential velocity $v_T$ for stars in the kinematic sample that are $\pm 0.5\,\kpc$ from the $x$-axis smoothed by a Gaussian kernel. The median velocity $v_T$ is calculated for $R_0 = 7.90\,\kpc$, $8.50\,\kpc$, and $\thisr\,\kpc$---our optimal value of $R_0$. All curves use $v_\odot/R_0 = 30.32\,\kmskpc$ from \citet{2019ApJ...885..131R}. The blue band shows the effect of varying $v_\odot/R_0$ within $8\sigma$ of the \citet{2019ApJ...885..131R} measurement, which translates to about a $9\,\kms$ variation in solar tangential velocity. For the optimal $R_0$, a minimum in $v_T$ at $\approx0\,\kms$ near the origin is observed $\approx 0.05\ \kpc$ away from $x=0$, well within the accuracy of our \texttt{astroNN} spectro-photometric distances. For non-optimal values of $R_0$, the minimum occurs away from $x=0$ and at larger values of $v_T$.}
\label{fig:vT_original}
\end{figure}

\figurename~\ref{fig:vT_original} displays $v_T$ as a function of $x$ for a few values of $R_0$. We see that for assumed values of $R_0$ away from the best-fit value, the minimum of the curve both shifts away from $x=0$ and the minimum has a higher $v_T$ value. Both of these indicate that these values of $R_0$ are disfavored. We also shows the effect of varying $v_\odot/R_0$ within its assumed uncertainty (eight times the nominal uncertainty from \citealt{2019ApJ...885..131R}) as a band around the curve for the best-fit value of $R_0$. It is clear that our solution is not very sensitive to the choice of $v_\odot/R_0$.

The kinematic maps of the disc and bar region using our optimal value of $R_0$ and $v_\odot/R_0=30.32\,\kmskpc$ from \citet{2019ApJ...885..131R} is shown in \figurename~\ref{fig:realdata_scenes}. In this figure, the left panel showing $v_T$ displays a minimum that is clearly shaped like a bar and the middle $v_R$ panel shows clear blue and red regions separated by the semi-major and semi-minor axes of the bar. Comparing these maps to those determined from the Milky-Way-like simulation in \figurename~\ref{fig:sims}, we see that they are remarkably similar.


\section{Discussion}\label{sec:discussions}

We have presented a new measurement of the distance $R_0$ to the Galactic center using the kinematics of stars in the Galactic bar. In this section, we compare our measurement to other recent measurements of $R_0$, we consider the implications for Sgr A$^*$, and we derive an updated measurement of the bar's pattern speed using the method of \citet{2019MNRAS.490.4740B} applied to the kinematic maps of \figurename~\ref{fig:realdata_scenes}.

\subsection{Comparison to other recent measurements of $R_0$}\label{subsec:litcomparison}

We compare our determination of $R_0 = \thisr\pm\thisrerr\,\kpc$ to a selection of other recent measurements in \figurename~\ref{fig:r0}. Visually, it is immediately clear that our measurement has a smaller uncertainty than essentially all but the S2-based measurements \citep{2018A&A...615L..15G,2019Sci...365..664D,2021A&A...647A..59G} (the \citealt{2013ApJ...769...88N} measurement being a notable exception that agrees well with ours). \citet{2016ARA&A..54..529B} summarized measurements up to $\approx 2016$ (thus excluding the recent, high-precision S2-based measurements) and derived an overall best estimate of $R_0 = 8.2 \pm 0.1\,\kpc$. This is in remarkably good agreement with our own measurement here both in value and uncertainty, but the value derived by \citet{2016ARA&A..54..529B} is a consensus value based on \emph{all} pre-2016 measurements, while ours is a measurement based on a single simple and robust signature of the Galactic center.

The most recent S2-based measurements using the GRAVITY instrument and using Keck have achieved high precision by taking advantage of S2's 2018 pericentric passage around Sgr A$^*$. The resulting precision is about a factor of 2 to 4 better than our measurement. However, all S2-based (or, more generally, Sgr A$^*$-based) measurements have to assume that Sgr A$^*$ is located at the center of the Milky Way. This does not have to be the case for a variety of reasons that are discussed in more detail in the next section: Sgr A$^*$ might not be at rest at the Galactic center even if it is orbiting the true center, or the nuclear star cluster (NSC) that contains Sgr A$^*$ might be sloshing around within the bar/disc system. In both cases, the offset between Sgr A$^*$ and the true barycenter can be larger than the current measurement uncertainties. Our measurement, on the other hand, is unambiguously of the barycenter of the bar/disc system, which are coupled because the bar is an integral dynamical part of the disc, and this is likely the actual barycenter. More prosaically, the S2-based methods are still somewhat plagued by instrumental effects like aberrations and by the difficulty of tying the small field-of-view reference frame to a global one. The difference between the \citet{2019Sci...365..664D} and \citet{2021A&A...647A..59G} measurements (and the significant differences between different iterations of the GRAVITY measurements based on similar data) demonstrate that these remain real issues and that the true uncertainty of the S2-based methods is likely larger than their reported uncertainty. Indeed, simply splitting the difference between \citealt{2019Sci...365..664D} and \citealt{2021A&A...647A..59G} as an estimate of their true uncertainty gives $R_0 \approx 8.15\pm 0.15\,\kpc$, consistent with our own measurement with similar uncertainty.

\subsection{Implications for the position and motion of Sgr A$^*$}\label{sec:implsgra}

As discussed above, recent measurements of $R_0$ based on the orbit of the star S2 around Sgr A$^*$ have achieved very high precision, but these measurements have to assume that Sgr A$^*$ is located at the barycenter of the Milky Way. This may not be the case. A supermassive black hole at the center of a dense NSC like the Milky Way's will perform Brownian motion resulting from the random combined kicks from passing stars \citep{2002ApJ...572..371C,2007AJ....133..553M}. The agreement between the value of $v_\odot/R_0 = 30.32\pm0.27\,\kmskpc$ obtained from the kinematics of masers \citep{2019ApJ...885..131R} and the value $v_\odot/R_0 = 30.39\pm0.04\,\kmskpc$ obtained from the proper motion of an assumed-at-rest Sgr A$^*$ \citep{2020ApJ...892...39R} demonstrates that the velocity of this Brownian motion is $\lesssim 0.27\,\kmskpc\times R_0 \approx 2\,\kms$. For the $\gtrsim 10^6\,M_\odot\,\pc^{-3}$ density of the NSC, this would lead to a spatial offset $\ll 1\,\pc$. Brownian motion of Sgr A$^*$ within the NSC on its own is therefore not a significant issue in determining $R_0$ using Sgr A$^*$.

Even if Sgr A$^*$ is at the center of the NSC, it is possible that the entire NSC is moving with respect to the barycenter of the entire Milky Way. The NSC may have formed through the accretion of a dense stellar system and in this case, the NSC sinks to the Galactic center through dynamical friction. However, if the stellar density is cored near the center---as direct measurements of the stellar density hint it might be within the central few hundred pc \citep{2013MNRAS.435.1874W}---the inspiraling NSC may stall near the core radius \citep{2016MNRAS.463..858P}. The NSC and Sgr A$^*$ may therefore find itself $\approx 100\,\pc$ away from the barycenter of the bar/disc. Observational determinations of the locations of supermassive black holes relative to the centers of their host galaxies indeed show that $\approx 100\,\pc$ offsets are common (see Fig. 1 and references in \citealt{2021PhRvD.103b3523B}). The agreement between non-Sgr-A$^*$-based and Sgr-A$^*$-based measurements of $v_\odot/R_0$ again limit the velocity of the motion of the NSC to be $\lesssim 2\,\kms$, implying that we should be seeing the NSC near a turning point if its orbit extends out to $\approx 100\,\pc$. While this is unlikely, it is within the realm of possibility. The agreement between our $R_0$ measurement and the Sgr-A$^*$-based ones shows that the offset between the NSC and the barycenter is $\lesssim 100\,\pc$, consistent with the previous considerations without providing a stringent constraint on them.  

\subsection{Pattern Speed}\label{subsec:patternspeed}

Because the Galactic bar dominates the mass distribution in the central regions of our Galaxy, determining its properties is essential to understanding the co-evolution of the bar and the other Galactic components and for studying the dynamics of stars in the bar, disc, and halo (e.g., \citealt{2017NatAs...1..633P,2018MNRAS.477.3945H,2018Natur.561..360A,2019MNRAS.484.2009B,2019MNRAS.488.3324F}). One of the most important properties that has been historically challenging to measure is the bar's pattern speed $\Omega_\mathrm{bar}$. Because the bar's pattern speed sets the locations of resonances in the disc and halo, its exact value is important for, e.g., the interpretation of the velocity structure near the Sun in terms of bar and spiral resonances (e.g., \citealt{2019MNRAS.490.1026H,2021MNRAS.500.2645T}. Measurements over the last decade have found a wide range of pattern speeds, with disc-kinematic measurements finding $\Omega_\mathrm{bar} \approx 50\,\kmskpc$ by assuming that the solar-neighbourhood Hercules stream is caused by the 2:1 outer Lindblad resonance (e.g., \citealt{2014A&A...563A..60A}), gas kinematics in the bar finding $\Omega_\mathrm{bar} \approx 33\,\kmskpc$ \citep{2016ApJ...824...13L}, and stellar kinematics preferring values at $\Omega_\mathrm{bar} \approx 40\,\kmskpc$ \cite{2017MNRAS.465.1621P,2019MNRAS.488.4552S,2019MNRAS.490.4740B}.

\citet{2019MNRAS.490.4740B} presented a purely-kinematic measurement of $\Omega_\mathrm{bar}$ based on the application of the continuity equation (similar to the famous \citealt{1984ApJ...282L...5T} method) to the kinematic maps of $v_T$ and $v_R$ in the bar region. They found $\Omega_\mathrm{bar}=41\pm3\,\kmskpc$ using similar data as we use here, but based on the earlier APOGEE DR16, \gaia\ DR2, and \texttt{astroNN} VAC DR16 data. For this measurement, they assumed a fiducial value of $R_0 = 8.125$ based on \citet{2018A&A...615L..15G} and $v_\odot=242\,\kms$. Repeating their measurement with the kinematic maps derived here that are shown in \figurename~\ref{fig:realdata_scenes} and using our best-fit value of $R_0=\thisr\,\kpc$ combined with  $v_\odot=249.44\,\kms$ (from combining $R_0$ with $v_\odot/R_0 = 30.32\,\kmskpc$), we find $\Omega_\mathrm{bar}=40.08\pm0.65\,\kmskpc$ where the uncertainty is purely statistical. This measurement is consistent with the measurement from \citet{2019MNRAS.490.4740B}, but slightly lower and with smaller statistical uncertainties (the statistical uncertainty in \citealt{2019MNRAS.490.4740B} was $1.5\,\kmskpc$).

Systematics in the measurement of $\Omega_\mathrm{bar}$ come in two flavours: model uncertainty about how well the method and its assumption work for a galaxy like the Milky Way and uncertainty in the adopted Galactic parameters $R_0$ and $v_\odot$. By applying the method to simulated galaxies that are realistic representations of the Milky Way's dynamics (including the simulation from \citealt{2017MNRAS.464..702K}), \citet{2019MNRAS.490.4740B} showed the model uncertainty to be $\approx 1.5\,\kms$ and this uncertainty remains. However, our improved measurement of $R_0$ leads to a decreased systematic uncertainty from the Galactic parameters.  \tablename\ \ref{table:pattern} gives the measured $\Omega_\mathrm{bar}$ for different sets of $R_0$ and $v_\odot$ to investigate this source of systematic uncertainty. The top section varies $R_0$ in the range $8.0$ to $8.4\,\kpc$ while keeping $v_\odot/R_0$ fixed to its fiducial value. In this case, $\Omega_\mathrm{bar}$ barely changes by $\approx 0.2\,\kmskpc$. The middle section shows what happens if we change $R_0$ or $v_\odot$ independently from their fiducial values by a large amount (compared to the uncertainties). In this case,  $\Omega_\mathrm{bar}$ changes significantly. Thus, the pattern speed is mainly affected by the solar tangential velocity, with only a minor effect from the uncertainty in $R_0$. The bottom section then more rigorously considers the effect of changing $v_\odot/R_0$ within its uncertainties from \citet{2019ApJ...885..131R}: when varying $v_\odot/R_0$ by $1\sigma$, $\Omega_\mathrm{bar}$ changes by $0.7\,\kmskpc$ (we also show the $8\sigma$ range considered previously to show the effect of large changes in $v_\odot/R_0$). Thus, the systematic uncertainty from the Galactic parameters is $\approx0.7\,\kmskpc$, about the same as the random uncertainty in the measurement. Note that we do not use the measurement of $v_\odot/R_0$ from \citet{2020ApJ...892...39R} that is derived from the proper motion of Sgr A$^*$ even though it has smaller uncertainties, because of the possibility of Brownian motion of Sgr A$^*$ (see the discussion in \secname~\ref{sec:implsgra} above).

Thus, our measurement of the pattern speed is
\begin{align}\label{eq:patspeed}
    \Omega_\mathrm{bar} & = 40.08\pm0.65\,(\mathrm{stat.})\,\pm 0.70\,(v_\odot/R_0\,\mathrm{unc.})\\
    & \ \ \quad \qquad \pm1.5\,(\mathrm{model\ unc.})\,\kmskpc\,,\nonumber\\
    & = 40.08\pm1.78\,\kmskpc\,,
\end{align}
where in the last line we have combined all the different sources of uncertainty into a single number.

\begin{center}
\begin{table}
\begin{tabular}{ |c|c|c|c| } 
\hline
$R_0\ (\kpc)$ & $v_\odot\ (\kms)$ & $\frac{v_\odot}{R_0}\ (\kmskpc)$ & $\Omega_{\mathrm{bar}}\ (\kmskpc)$ \\
\hline
\thisr & 249.44 & 30.32 & $40.08\pm0.65$ \\ 
8.00 & 242.56 & 30.32 & $39.90\pm0.75$  \\ 
8.40 & 254.69 & 30.32 & $39.95\pm0.61$ \\ 
\hline
8.00 & 249.44 & 31.18 & $42.00\pm0.71$ \\ 
\thisr & 240.00 & 29.17 & $37.25\pm0.64$ \\
\hline
\thisr & 247.22 & $30.32-0.27$ & $39.36\pm0.63$  \\ 
\thisr & 251.66 & $30.32+0.27$ & $40.78\pm0.65$  \\
\thisr & 231.67 & $30.32-2.16$ & $34.85\pm0.69$  \\ 
\thisr & 267.21 & $30.32+2.16$ & $44.12\pm0.69$  \\
\hline
\end{tabular}
\caption{Pattern speed $\Omega_\mathrm{bar}$ measurement using different sets of $R_0$ and $v_\odot$. The first row uses the best-fit $R_0$ from this paper combined with the $v_\odot/R_0$ measurement from \citet{2019ApJ...885..131R}. The top section shows the result of changing $R_0$ while keeping $v_\odot/R_0$ fixed. The middle section varies $R_0$ and $v_\odot$ by significant amount from their top-row values to demonstrate the effect of varying them independently. The bottom section shows how $\Omega_{bar}$ changes when varying $v_\odot/R_0$ by $1\sigma$ (first two rows of the section) or by $8\sigma$ (last two rows). The pattern speed is largely insensitive to changes in $R_0$ that keep $v_\odot/R_0$ constant, but is sensitive to variations in $v_\odot / R_0$.}
\label{table:pattern}
\end{table}
\end{center}

\section{Conclusion}\label{sec:conclusion}

The distance $R_0$ between the Sun and the Milky Way's barycenter is one of the most important parameters of the Milky Way, yet few previous measurements clearly and explicitly determine the distance to the barycenter itself, rather than to the region of highest stellar density or to the Milky Way's supermassive black hole. While it is likely that all three definitions of the center coincide in practice, they do not have to. In this paper, we have presented a direct, simple, and robust measurement of the distance to the barycenter of the bar+disc system based on the kinematics of stars in the Galactic bar region. The barycenter of the bar+disc likely coincides with the overall barycenter, unless the dark-matter halo sloshes around with respect to the disc, which is unlikely given the relatively quiescent recent history of the Milky Way. Our measurement simply searches for the expected minimum in the rotational velocity $v_T$ of stars along the Sun--Galactic-center line and determines $R_0$ by placing this minimum at the Galactic center. Our resulting measurement is
\begin{equation}
    R_0 = \thisr\pm\thisrerr\,\kpc\,.
\end{equation}
Our measurement is based on spectro-photometric distances, obtained from using the \texttt{astroNN} method to APOGEE DR17 spectra, combined with kinematics from APOGEE and \gaia\ EDR3. Extensive tests of our distance accuracy demonstrates that systematics in the distances are $\lesssim 1\%$, such that distance systematics do not significantly contribute to the uncertainty.

We have also used the kinematics of stars in the bulge to perform an updated measurement of the bar's pattern speed of $\Omega_\mathrm{bar} = 40.08\pm1.78\,\kmskpc$ using the method of \citet{2019MNRAS.490.4740B}. Equation \eqref{eq:patspeed} gives a detailed accounting of the different sources of random and systematic uncertainty in the error budget. This value is in good agreement with other recent measurements, which mostly agree that $\Omega_\mathrm{bar} \approx 40\,\kmskpc$ (e.g., \citealt{2017MNRAS.465.1621P,2019MNRAS.488.4552S,2019MNRAS.490.4740B,2022ApJ...925...71L}). 

Our measurement of $R_0$ is more precise and more accurate than most previous measurements, while also being one of the most simple and direct. This improvement over previous measurements is the result of (a) a significant improvement in the quality and coverage of stellar kinematics in the Galactic bar by combining the APOGEE and \gaia\ data and (b) the ability of modern machine-learning methods such as \texttt{astroNN} \citep{2019MNRAS.489.2079L} and other methods (e.g., \citealt{2019AJ....158..147H}) to leverage the precise \gaia\ parallaxes and create high-precision, high-accuracy spectro-photometric distances for large samples of stars at the distance of the Galactic center. The \texttt{astroNN} distances have a precision of $\approx 5\%$ with systematics $\lesssim 1\%$; this is crucial for attaining our high-precision measurement of $R_0$.

The only class of measurements that have smaller uncertainties than ours are measurements based on the orbit of the star S2 around Sgr A$^*$, the Milky Way's supermassive black hole. The best such measurements have uncertainties that are a factor of a few smaller than ours (e.g., \citealt{2019Sci...365..664D,2021A&A...647A..59G}). However, these measurements are not direct determinations of the distance to the Milky Way's barycenter, but have to make the additional assumption that Sgr A$^*$ is at rest at the barycenter. While this is a plausible assumption, it does not have to be the case (see \secname~\ref{sec:implsgra}). The S2-based methods are also still plagued by instrumental effects like aberrations and by the difficulty of tying the small field-of-view reference frame to a global one, while measurements such as ours do not suffer from the same systematics.

Future data will allow improved measurements of the type performed in this paper. First, the ongoing SDSS-V Milky Way Maper survey will increase the number of stars in the bar region by a factor of 100, significantly increasing the size of the sample where we have data now and increasing the bar's coverage to include the entire bar. While we will still have to rely on spectro-photometric distances with this sample, the full coverage of the bar will allow the use of the full two-dimensional kinematic maps to determine $R_0$, the pattern speed, and the detailed orbit distribution in the bar. Further in the future, Small-Jasmine will perform an astrometric survey of the inner Galaxy ($l \approx b \approx 0^\circ$) with an accuracy of $25\,\muas$ in parallax and $25\,\muas\,\mathrm{yr}^{-1}$ in proper motion \citep{2020IAUS..353...51G}. Because the minimum in $v_T$ that we use to determine $R_0$ is essentially a signature in proper motion $\mu_l$, Small-Jasmine will be able to measure $R_0$ using a similar method as ours, but using parallax distances instead. Similarly, the bulge microlensing survey that will be carried out by the Roman telescope will allow for $10\%$ parallaxes and $10\muas\,\mathrm{yr}^{-1}$ proper-motion measurements for millions of disk and bar stars in a $\approx 2\,\mathrm{deg}^2$ region towards the Galactic center \citep{2019BAAS...51c.211G}. A subset of $10^6$ giants of these will have ultra-precise ($\approx 0.3\%$) parallaxes \citep{2015JKAS...48...93G}. Using our method, these parallaxes and proper motions can lead to a high-precision, distance-systematics-free measurement of the distance $R_0$ to the Milky Way's barycenter.

\section*{Acknowledgements}

HL and JB acknowledge financial support from NSERC (funding reference number RGPIN-2020-04712) and an Ontario Early Researcher Award (ER16-12-061).

Funding for the Sloan Digital Sky 
Survey IV has been provided by the 
Alfred P. Sloan Foundation, the U.S. 
Department of Energy Office of 
Science, and the Participating 
Institutions. 

SDSS-IV acknowledges support and 
resources from the Center for High 
Performance Computing  at the 
University of Utah. The SDSS 
website is www.sdss.org.

SDSS-IV is managed by the 
Astrophysical Research Consortium 
for the Participating Institutions 
of the SDSS Collaboration including 
the Brazilian Participation Group, 
the Carnegie Institution for Science, 
Carnegie Mellon University, Center for 
Astrophysics | Harvard \& 
Smithsonian, the Chilean Participation 
Group, the French Participation Group, 
Instituto de Astrof\'isica de 
Canarias, The Johns Hopkins 
University, Kavli Institute for the 
Physics and Mathematics of the 
Universe (IPMU) / University of 
Tokyo, the Korean Participation Group, 
Lawrence Berkeley National Laboratory, 
Leibniz Institut f\"ur Astrophysik 
Potsdam (AIP),  Max-Planck-Institut 
f\"ur Astronomie (MPIA Heidelberg), 
Max-Planck-Institut f\"ur 
Astrophysik (MPA Garching), 
Max-Planck-Institut f\"ur 
Extraterrestrische Physik (MPE), 
National Astronomical Observatories of 
China, New Mexico State University, 
New York University, University of 
Notre Dame, Observat\'ario 
Nacional / MCTI, The Ohio State 
University, Pennsylvania State 
University, Shanghai 
Astronomical Observatory, United 
Kingdom Participation Group, 
Universidad Nacional Aut\'onoma 
de M\'exico, University of Arizona, 
University of Colorado Boulder, 
University of Oxford, University of 
Portsmouth, University of Utah, 
University of Virginia, University 
of Washington, University of 
Wisconsin, Vanderbilt University, 
and Yale University.

This work has made use of data from the European Space Agency (ESA) mission
{\it Gaia} (\url{https://www.cosmos.esa.int/gaia}), processed by the {\it Gaia}
Data Processing and Analysis Consortium (DPAC,
\url{https://www.cosmos.esa.int/web/gaia/dpac/consortium}). Funding for the DPAC
has been provided by national institutions, in particular the institutions
participating in the {\it Gaia} Multilateral Agreement.

\section*{Data Availability}

The basic data described in \secname~\ref{sec:data} that this work is based on is available in the \texttt{astroNN} DR17 VAC available as part of SDSS-IV's DR17. All of the code underlying this article is available on GitHub at \url{https://github.com/henrysky/astroNN_GC_distance}



\bibliographystyle{mnras}
\bibliography{mnras_template}







\bsp	
\label{lastpage}
\end{document}